\documentclass[12pt]{article} 
\usepackage[numbers]{natbib}
\usepackage{a4}
\usepackage{amsmath}
\usepackage{amssymb}
\usepackage{latexsym}
\usepackage{amssymb}
\usepackage{epsfig}
\usepackage{natbib}
\def \TP{{\mathrm{P}}}

\def \pd{\partial}

\def \e{{\mathrm{e}}}

\def \Bbeta{\boldsymbol{\beta}}

\def \Bu{{\boldsymbol{u}}}

\def \Bx{{\boldsymbol{x}}}

\def \BR{{\boldsymbol{R}}}

\def \BJ{{\boldsymbol{J}}}
\def \BB{{\boldsymbol{B}}}
\def \BA{{\boldsymbol{A}}}

\def \Bbeta{\boldsymbol{\beta}}
\def \Balpha{\boldsymbol{\alpha}}

\def \Bu{{\boldsymbol{u}}}

\begin{document}
%%%%%%%%%%%%%%%%%%%%%%%%%%%%%%%%%%%%%%%%%%%%%%%%%%%%%%%%%%%%%%%%%%%%%
\title{{\bf On gradient field theories:\\ 
gradient magnetostatics and gradient elasticity
}}
\author{
Markus Lazar~$^\text{}$\footnote{%%%Corresponding author. 
{\it E-mail address:} lazar@fkp.tu-darmstadt.de (M.~Lazar).} 
\\ \\
${}^\text{}$ 
        Heisenberg Research Group,\\
        Department of Physics,\\
        Darmstadt University of Technology,\\
        Hochschulstr. 6,\\      
        D-64289 Darmstadt, Germany
%%%${}^\text{b}$ 
}

\date{\today}    
\maketitle
%%%%%%%%%%%%%%%%%%%%%%%%%%%%%%%%%%%%%%%%%%%%%%%%%%%%%%%%%%%%%%%%%%%%%%%%%%%%%

\begin{abstract}
In this work the fundamentals of gradient field theories 
are presented and reviewed. 
In particular, the theories of gradient magnetostatics and
gradient elasticity are investigated and compared. 
For gradient magnetostatics, 
non-singular expressions for
the magnetic vector gauge potential, the Biot-Savart law, the Lorentz force and the 
mutual interaction energy of two electric current loops are derived and discussed.
For gradient elasticity, non-singular forms of all 
dislocation key-formulas (Burgers equation, Mura equation, Peach-Koehler
stress equation, Peach-Koehler force equation, and
mutual interaction energy of two dislocation loops) are presented.
In addition, similarities between an electric current loop and 
a dislocation loop are pointed out.
The obtained fields for both gradient theories are non-singular due to a 
straightforward and self-consistent 
regularization.
\\

\end{abstract}
%%%\vspace{1cm}

\newpage

\section{Introduction}

Nowadays, gradient theories are very popular in physics, applied mathematics, material science
and engineering science. 
Gradient theories are theories possessing internal length scales 
in order to describe size-effects. 
Such theories provide non-singular solutions of the field equations and
therefore a regularization is achieved 
(e.g. a dislocation core regularization and an electron core regularization of the 
classical Dirac delta expressions).
The principle concept of a gradient theory is simple;
in addition to the ``classical'' state quantities, their gradient terms also have to be 
implemented in the Lagrangian density (or energy density). 
For a gradient theory of order $n$, the Lagrangian density depends on 
all the gradients of the state quantities up to order $n$.
From that point of view, gradient theories are effective theories.

In physics, the most popular gradient theory is the so-called Bopp-Podolsky theory,
which is the gradient version of the theory of electrodynamics. 
\citet{Bopp} and \citet{Podolsky} 
have proposed theories representing generalizations of the 
theory of electrodynamics to linear field equations of fourth-order 
in order to avoid singularities in electrodynamics
(see also~\citep{PS,Iwan,Davis}).
Such a generalized electrodynamics has a physical meaning 
if the (static) electric potential becomes the Coulomb potential asymptotically, 
and if the point-like field sources have a Dirac delta form~\citep{Treder}.
The Bopp-Podolsky theory has many interesting features.
It solves the problem of infinite energy in the electrostatic case, and 
it gives the correct expression for the self-force of charged particles at 
short distances eliminating the singularity when $r\rightarrow 0$
as shown by ~\citet{Frenkel}. 
In this manner, the Bopp-Podolsky electrodynamics is free of divergences. 
Another important prediction of the Bopp-Podolsky theory is that 
the value of an electron core radius is proportional to 
a parameter $a$, the so-called Bopp-Podolsky parameter.
These features allow experiments that could test the generalized
electrodynamics as a viable effective field theory~(e.g., \citep{Cuzi}).
\citet{Iwan} and \citet{Kvasnica} argued that the Bopp-Podolsky parameter $a$ 
is in the order of $\sim 10^{-13}$\, cm.
From a historical point of view, it is mentioned that in the sixties Feynman~\citep{Feynman}
has never appreciated the usefulness and power of the Bopp-Podolsky theory. 
The sixties were a time when physicists were usually more interested in quantum theories.
However, the Bopp-Podolsky model has a close relationship with the Pauli-Villars 
regularization procedure used in quantum electrodynamics 
(see, e.g.,~\citep{Kvasnica2,Zayats}). 
In this way, the Bopp-Podolsky theory serves a ``physical'' regularization method based on  
higher-order partial differential equations.

It should be mentioned that one can find 
in the literature (e.g.,~\citep{Maugin12}) 
that the classical theory of Maxwell's electrodynamics 
and Einstein's theory of gravity (general relativity) are gradient theories.
However, both theories do not possess characteristic length scales.
No classical theories exist 
such that  the theories of electrodynamics and of 
general relativity could be considered as their gradient version.
Moreover, the electromagnetic potentials are gauge
fields and are not gauge invariant, and therefore they are not state quantities. 
Only the electromagnetic field strengths are  state quantities, 
reiterating Maxwell's theory of ``classical'' electrodynamics,
and therefore the corresponding gradient theory is the Bopp-Podolsky theory.

More than twenty years after the Bopp-Podolsky theory, 
\citet{Mindlin64} (see also~\citep{Mindlin68,Mindlin72,Jaunzemis,Eshel,Eshel73})
introduced theories of gradient elasticity however without giving any credit to Bopp and Podolsky.
In order to remove the singularities in the classical solutions, such continuum theories of 
generalized elasticity may be used. 
The correspondence between the gradient elasticity theory and the atomic structure
of materials with the nearest and next nearest interatomic interactions was
exhibited by~\citet{Toupin}.
The original Mindlin theory~\citep{Mindlin64,Mindlin68} possesses many additional material parameters.
For isotropic materials,
Mindlin's theory of first strain gradient elasticity~\citep{Mindlin64,Mindlin68}
possesses two characteristic lengths.
The discrete nature of materials is inherently incorporated in the formulation
through the characteristic lengths. 
One can say that gradient elasticity is a continuum theory valid on small scales.
The capability of strain gradient theories in capturing size effects
is a direct manifestation of the involvement of characteristic lengths. 
\citet{Lardner71} was the first who investigated 
straight screw and edge dislocations in the framework of 
Mindlin's gradient elasticity theory. 
Since considering neither plastic distortion nor dislocation density,
\citet{Lardner71} constructed actually solutions 
for  a compatible boundary value problem,
which are still singular.

Simplified versions, which are  particular cases of 
Mindlin's theories, have been proposed and used in the literature.
A simplified gradient elasticity theory 
is gradient elasticity of Helmholtz type~\citep{LM05,LMA05,Lazar13}, with only one material 
length scale parameter as new material coefficient. 
The theory of gradient elasticity of Helmholtz type is a 
special  version of Mindlin's gradient elasticity theory~\citep{Mindlin64}.
Using ab initio calculations, 
\citet{Shodja} found that the characteristic length scale parameters
of first strain gradient elasticity 
are of the order  $\ell\sim 10^{-10}$~m for several fcc and bcc materials.
Therefore, gradient elasticity can be used for understanding the nano-mechanical
phenomena at such length scales.

Non-singular fields of straight dislocations and dislocation loops
were obtained in the framework of gradient elasticity of Helmholtz type
by~\citet{LM05,LM06} and \citet{Lazar12,Lazar13}, respectively.
\citet{Lazar12,Lazar13} derived the non-singular 
dislocation key-formulas (Burgers formula, Mura formula
and Peach-Koehler stress formula) valid in gradient elasticity.
Such non-singular solutions of arbitrary dislocations
might be very useful for the so-called discrete dislocation dynamics (e.g.,~\citep{Li,Sun}). 
Since dislocations are the basic carriers of plasticity,  
the fundamental physics of plastic deformation must be described in terms of
the behavior of dislocation ensembles.
\citet{LM06} have shown that, for straight dislocations, 
the gradient parameter leads to a smoothing of the displacement profile, 
in contrast to the jump occurring in the classical solution.
\citet{Lazar07} has extended gradient elasticity of Helmholtz type 
for functionally graded materials and an analytical solution of a screw dislocation 
in such a material was given.

Gradient elasticity with only one gradient parameter
can be found in the literature also under the names dipolar gradient elasticity theory~\citep{G03,Georg04}, 
simplified strain gradient elasticity theory~\citep{GM10a,GM10b}
and special gradient elasticity theory~\citep{AA97}. 
Useful applications of such a gradient elasticity theory are for example
cracks (e.g., \citep{G03}) as well as the Eshelby inclusion problem
(e.g., \citep{GM10a,GM10b}).  
However, the framework of \citet{AA97} and \citet{GA99,GA99b} lacks double stresses
and is not based on proper variational considerations (e.g. to obtain pertinent 
boundary conditions).
It is remarkable that~\citet{Guenther76} was
the first who spoke of a mechanical model of the Bopp-Podolsky potential for 
defects in elasticity.
Like the Bopp-Podolsky theory, 
gradient elasticity theory of Helmholtz type serves a ``physical''
regularization based on  higher-order partial differential equations.
A nice overview on gradient theories in physics
(superconductivity, radiative fluid dynamics, theory of dielectrics, and surface phenomena)
was given by~\citet{Maugin79}.

The aim of this paper is to present a comparison 
between the magnetostatic Bopp-Podolsky theory 
and the theory of gradient elasticity of Helmholtz type.
Similarities and differences for these theories are pointed out.
In addition, we derive new key-equations for both gradient theories.
For electric current loops, the Biot-Savart law, 
the Lorentz force, and the mutual interaction energy are derived, for the first time, 
in the framework of gradient magnetostatics. 
For dislocation loops, all the dislocation key-formulas
(Burgers equation, Mura equation, Peach-Koehler stress equation, Peach-Koehler force
equation, mutual interaction energy) are calculated using gradient elasticity.
Moreover, following the analogy between ``classical'' magnetostatics and ``classical''
dislocation theory pointed out by~\citet{deWit60}, we investigate the 
analogy between gradient magnetostatics and dislocations in gradient elasticity.
We consider in both theories an infinite continuum, therefore there is no need
for boundary conditions. 
For completeness, boundary conditions are given in the Appendix~B.
Moreover, we decompose the boundary conditions into the classical part and
the gradient part and we also give a physical interpretation of them.

The paper is organized as follows. 
In Section~2, the fundamentals of gradient theory of magnetostatics are
presented and the Biot-Savart law, 
the Lorentz force, and the mutual interaction energy are calculated.
The ``Bifield'' ansatz 
for gradient magnetostatics~\citep{Bopp,Podolsky} is used 
for the decomposition of magnetic fields into the classical part and a purely 
gradient part.
In Section~3, the theory of gradient elasticity of Helmholtz type is reviewed
and investigated. 
Dislocations are examined in the framework of gradient elasticity. 
A ``Ru-Aifantis theorem'' is generalized for dislocations in 
an infinite medium in the framework of gradient elasticity of Helmholtz type. 
In addition, a ``Bifield'' ansatz for gradient elasticity is introduced
and used for the decomposition of fields into the classical part and a purely 
gradient part.
All dislocation key-formulas valid in gradient elasticity are given.
The presentation of the two gradient theories reveals the similarities and
differences between them.
In Section~4, the final conclusions are given. 
Some mathematical and technical details and a discussion of the boundary conditions 
are presented in the Appendices.

\section{Gradient magnetostatics -- Bopp-Podolsky theory}
In this section, we investigate the gradient theory of 
magnetostatics which is the magnetostatic part of the
Bopp-Podolsky theory \citep{Bopp,Podolsky}.
In such a theory of gradient magnetostatics, the energy density takes the form
\begin{align}
\label{W-BP}
W=\frac{1}{2\mu_0}\, B_{i} B_{i}+\frac{1}{2\mu_0}\, a^2 \pd_k B_{i} \pd_k B_{i}
-A_k J_k\, ,
\end{align}
where $B_i$ denotes the magnetic field vector (or magnetic induction), 
$\mu_0$ is the permeability of vacuum,
$a$ is taken to be 
a fundamental constant with dimension of length, $J_k$ is the electric 
current density vector
and $A_k$ denotes the magnetic vector gauge potential.
The magnetic field vector may be expressed in terms of the 
magnetic vector gauge potential
\begin{align}
\label{B-A}
B_i=\epsilon_{ijk}\pd_j A_k\,,
\end{align}
satisfying the Bianchi identify
\begin{align}
\label{BI1}
\pd_i B_i=0\,,
\end{align}
which means that magnetic monopoles do not exist.
Here, $\epsilon_{ikl}$ denotes the Levi-Civita tensor.
%%%%It can be seen that the energy density~(\ref{W-BP}) 
%%%%exhibits a  symmetry both in 
%%%$H_{i}$ and $B_{i}$ and in $\pd_k H_{i}$ and $\pd_k B_{i}$.

From Eq.~(\ref{W-BP}), two kinds of excitation fields can be defined
\begin{align}
\label{H}
H_i &=\frac{\pd W}{\pd B_i}=\mu_0^{-1}\, B_i\,,\\
\label{H2}
H_{ik} &=\frac{\pd W}{\pd (\pd_k B_i)}=a^2 \mu_0^{-1}\, \pd_k B_i
=a^2 \pd_k H_i\,,
\end{align} 
where 
$H_i$ is the magnetic excitation vector and 
$H_{ik}$ is the magnetic excitation tensor, which is a higher-order excitation field.
It can be seen in Eq.~(\ref{H2}) that $H_{ik}$ is just the gradient 
of $H_{i}$ and multiplied by $a^2$. 
From Eqs.~(\ref{BI1})--(\ref{H2}), 
it follows: $\pd_i H_i=0$ and $\pd_i H_{ik}=0$.
In addition, it yields
\begin{align}
\frac{\pd^2 W}{\pd B_{i}\, \pd B_{i}}=\frac{1}{\mu_0}\,,\qquad\quad
\frac{\pd^2 W}{\pd (\pd_k B_{i})\, \pd (\pd_k B_{i})}
=\frac{a^2}{\mu_0}\,.
\end{align}

Using a variational principle with respect to the magnetic vector gauge potential $A_i$, 
the Euler-Lagrange equation is given by (e.g., \citep{Iwan})
\begin{align}
\label{EL-A}
\frac{\delta W}{\delta A_i}=\frac{\pd W}{\pd A_i}
-\pd_j\, \frac{\pd W}{\pd (\pd_j A_i)}
+\pd_k\pd_j\,  \frac{\pd W}{\pd (\pd_k \pd_j A_i)}=0\,.
\end{align}
By means of Eqs.~(\ref{W-BP}), (\ref{H}) and (\ref{H2}), the Euler-Lagrange equation~(\ref{EL-A}) reduces to 
\begin{align}
\label{FE-H0}
\epsilon_{ijk}\pd_j \big(H_k-\pd_l H_{kl}\big)=J_i\,.
\end{align}
Using Eq.~(\ref{H2}), Eq.~(\ref{FE-H0}) can be simplified to
\begin{align}
\label{FE-H}
L\epsilon_{ijk}\pd_j H_k=J_i
\end{align}
with the Helmholtz operator $L$ depending on the length scale $a$
\begin{align}
L=1-a^2\Delta\,,
\end{align}
where $\Delta=\pd_i\pd_i$ denotes the Laplacian.
Eqs.~(\ref{FE-H0}) and (\ref{FE-H}) are the Amp\`ere law valid in gradient magnetostatics.
Eqs.~(\ref{BI1}) and (\ref{FE-H}) are the field equations for gradient 
magnetostatics.
The field equation~(\ref{FE-H}) is a partial differential equation (pde) of 3rd-order 
for the field\footnote{A more general constitutive relation than Eq.~(\ref{H2}) is
$H_{ik}=c_1 \pd_k B_i+c_2 \pd_i B_k$, since $\delta_{ik} \pd_l B_l=0$. 
Using Eq.~(\ref{BI1}), it does not change the Euler-Lagrange equation~(\ref{FE-H0}), due to
$\pd_l H_{kl}=c_1 \Delta B_k$, and $c_1=a^2 \mu_0^{-1}$.
Therefore, gradient magnetostatics possesses in a natural way only one internal
length scale parameter, namely $a$.} $H_k$. 
In addition, the current vector satisfies the continuity equation
\begin{align}
\label{DivJ}
\pd_i J_i=0\,.
\end{align}

Taking the curl of Eq.~(\ref{FE-H}) and using $\pd_i H_i=0$, 
it can be written in the form 
of an inhomogeneous Helmholtz-Laplace equation (pde of 4th-order)
\begin{align}
\label{FE-H2}
L \Delta H_i=-\epsilon_{ijk}\pd_j J_k\,.
\end{align}
Using Eqs.~(\ref{B-A}) and (\ref{H}), Eq.~(\ref{FE-H}) reduces to a
field equation for the magnetic vector gauge potential (pde of 4th-order)
\begin{align}
L (\pd_i\pd_k -\delta_{ik}\Delta)A_k=\mu_0\, J_i\,.
\end{align}

If the Coulomb gauge  condition, which is a side condition, 
is used for the magnetic vector gauge potential $A_k$,
\begin{align}
\label{CG}
\pd_k A_k=0\,,
\end{align}
or the generalized Coulomb gauge  condition
\begin{align}
\label{CG1}
L \pd_k A_k=0\,,
\end{align}
then the magnetic vector gauge potential $A_k$ satisfies the
following inhomogeneous Helmholtz-Laplace equation 
which is a pde of 4th-order for $A_k$
\begin{align}
\label{HL-A}
L \Delta A_k=-\mu_0\, J_k\,.
\end{align}
The formal solution of Eq.~(\ref{HL-A}) is given as convolution
\begin{align}
\label{A-conv}
A_k=-\mu_0\, G*J_k\,,
\end{align}
where $*$ denotes the spatial convolution and
$G$ denotes here the Green function of the Helmholtz-Laplace equation and is
defined by
\begin{align}
\label{G-pde}
L \Delta G=\delta(\Bx-\Bx')\,.
\end{align}
The three-dimensional solution of the Green function of the Helmholtz-Laplace equation reads
\begin{align}
\label{G-LH}
G(R)=-\frac{1}{4\pi R}\, \Big(1-\e^{-R/a}\Big)\,,
\end{align}
where $R=|\Bx-\Bx'|$. 
Eq.~(\ref{G-LH}) represents the regularized Green function in the static Bopp-Podolsky theory and $G(0)=-1/[4\pi a]$.
Using Eq.~(\ref{A-conv}) and 
the property of the differentiation of a convolution~\citep{Wl,Kanwal}, 
it can be seen that the Coulomb gauge condition~(\ref{CG}) is fulfilled 
as a consequence of the continuity equation~(\ref{DivJ})
\begin{align}
\label{CG2}
\pd_k A_k=-\mu_0\,\pd_k (G*J_k)=-\mu_0\, G*(\pd_k J_k)=0\,.
\end{align}

The substitution of Eq.~(\ref{G-LH}) into Eq.~(\ref{A-conv}) gives the 
solution for the magnetic vector gauge potential
\begin{align}
\label{A-int}
A_k=\frac{\mu_0}{4\pi}\int_{V} \frac{1}{R}\, \Big(1-\e^{-R/a}\Big)\,J_k(\Bx')\, d V'\,,
\end{align}
which vanishes at infinity. 
Using Eqs.~(\ref{B-A}) and (\ref{A-int}), the magnetic field vector is calculated as
\begin{align}
\label{B-int}
B_i&=-\frac{\mu_0}{4\pi}\,\int_{V}\epsilon_{ijk}\,\frac{R_j}{R^3} 
\bigg[1-\bigg(1+\frac{R}{a}\bigg)\e^{-R/a}\bigg]\, J_k(\Bx')\, d V'\,,
\end{align}
which is the general Biot-Savart law for a volume current $J_k$ 
valid in gradient magnetostatics. 
Eq.~(\ref{B-int}) determines the non-singular magnetostatic field 
of a current distribution $ J_k(\Bx')$.
Here, $\BR=\Bx-\Bx'$ denotes the relative radius vector.
In the limit $a\rightarrow 0$, Eqs.~(\ref{A-int}) and (\ref{B-int})
reduce to the ``classical'' results of magnetostatics (see, e.g., \citep{Jackson,PP}).
The fields~(\ref{A-int}) and (\ref{B-int}) are non-singular.

If $J_k$ is the ``true'' electric current,  we may 
introduce a so-called ``free'' electric current $J_k'$ (or ``effective'' 
electric current) by 
\begin{align}
\label{J-true}
J_k=L J_k'\,.
\end{align}
Then the field equations~(\ref{FE-H2}) and (\ref{HL-A}) are modified to
\begin{align}
\label{FE-H-true}
L \Delta H_i&=-\epsilon_{ijk}\pd_j LJ'_k\,,\\
\label{HL-A-true}
L \Delta A_k&=-\mu_0\, L J_k'\,.
\end{align}

Alternatively, the field equation~(\ref{FE-H}), 
which is a pde of 3rd-order, may be rewritten
as an analogous system of pdes, namely one of 1st-order and another one
of 2nd-order,
\begin{align}
\label{RA-H0}
\epsilon_{ijk}\pd_j H^0_k&=J_i\,,\\
\label{RA-H1}
L H_k&=H_k^0\,.
\end{align}
In addition, it yields
\begin{align}
H^0_i =\mu_0^{-1}\, B^0_i\,,
\end{align} 
and
\begin{align}
B^0_i=\epsilon_{ijk}\pd_j A^0_k\,.
\end{align}
The corresponding Bianchi identity reads
\begin{align}
\label{BI0}
\pd_i B^0_i=0\,.
\end{align}

\subsection{Bifield-Ansatz}

Since the Bopp-Podolsky theory is a generalization of the classical
electrodynamics,
the question arises how the classical fields can be separated from the
generalized fields.
The considered type of linear theory possesses the interesting property that 
the field $A_k$ might be represented as a superposition of two other fields
(the so-called ``Bifield'') 
\begin{align}
\label{A-deco}
A_k=A_k^0+A_k^1\,,
\end{align}
satisfying the following equations of second-order (e.g.,~\citep{Iwan,Davis}).
Since $A_k^0$ satisfies an inhomogeneous Laplace equation (or Poisson equation)
\begin{align}
\label{A0-pde}
\Delta A^0_k=-\mu_0\, J_k\,,
\end{align}
$A_k^0$ may be identified with the classical magnetic vector gauge potential.
$A_k^1$ is the part of the  magnetic gauge potential depending on
the parameter $a$ and, therefore, it is called the gradient part.
In addition, $A_k$ fulfills the inhomogeneous Helmholtz equation
\begin{align}
\label{LA}
L A_k=A^0_k\,,
\end{align}
and the Poisson equation
\begin{align}
\label{PA}
a^2\Delta A_k=A^1_k\,.
\end{align}
%%%where $A_k^0$ is the source term. 
Substituting Eq.~(\ref{A-deco}) into Eq.~(\ref{LA}) and using Eq.~(\ref{A0-pde}),
we obtain for the gradient part $A_k^1$ the following equation
\begin{align}
L A^1_k=a^2 \Delta A^0_k=-\mu_0\, a^2 J_k\,.
\end{align}
Thus, the field $A_k^0$ satisfies an inhomogeneous Laplace equation
and the field $A_k^1$ satisfies an inhomogeneous Helmholtz equation.
In both cases, the source field is $J_k$. 
Using Eq.~(\ref{LA}), the generalized Coulomb gauge condition~(\ref{CG1})
reduces to the Coulomb gauge condition for $A_k^0$
\begin{align}
\label{CG0}
L \pd_k A_k=\pd_k A_k^0=0\,.
\end{align}

Also for the magnetic excitation vector field $H_k$, we may make a ``Bifield'' ansatz:
\begin{align}
H_k=H_k^0+H_k^1\,,
\end{align}
where $H_k^0$ fulfills the following Poisson equation 
\begin{align}
\label{FE-H-P}
\Delta H_k^0=-\epsilon_{kji}\pd_j J_i\,.
\end{align}
In addition to Eqs.~(\ref{RA-H0}) and (\ref{RA-H1}), the 
following equations hold
\begin{align}
L H_k^1&=a^2\Delta H_k^0=-a^2 \epsilon_{kji}\pd_j J_i\,,\\
\label{PH}
a^2\Delta H_k&=H^1_k\,,
\end{align}
as well as 
\begin{align}
H_k^1=\pd_i H_{ki}\,.
\end{align}

A ``Bifield'' ansatz for the magnetic field vector $B_k$ is given by
\begin{align}
\label{B-deco}
B_k=B_k^0+B_k^1\,,
\end{align}
where $B_k^0$ satisfies the following Poisson equation 
\begin{align}
\label{FE-H20}
\Delta B_k^0=-\mu_0 \epsilon_{kji}\pd_j J_i
\end{align}
and the equations
\begin{align}
\label{RA-B0}
\epsilon_{ijk}\pd_j B^0_k&=\mu_0\, J_i\,,\\
\label{RA-B1}
L B_k&=B_k^0\,,
\end{align}
as well as 
\begin{align}
\label{LB1}
L B_k^1&=a^2\Delta B_k^0=-\mu_0 a^2 \epsilon_{kji}\pd_j J_i\,,\\
\label{PB}
a^2\Delta B_k&=B^1_k\,.
\end{align}

Using the ``Bifield''-ansatz, the regularized Green function~(\ref{G-LH})
can be decomposed into two parts 
\begin{align}
\label{G-deco}
G=G^0+G^1\,,
\end{align}
where
\begin{align}
\label{G-bi}
G^0=-\frac{1}{4\pi R}\,,\qquad 
G^1=\frac{1}{4\pi R}\,\e^{-R/a}\,,
\end{align}
satisfying the following equations of second-order 
\begin{align}
\label{G0-pde}
\Delta G^0=\delta(\Bx-\Bx')\,.
\end{align}
Here $G^0$ is the Green function of the Laplace operator and
$G^1$ is the Green function of the Helmholtz operator.
In addition, $G$ fulfills the inhomogeneous Helmholtz equation
\begin{align}
\label{LG}
L G=G^0\,,
\end{align}
and the Poisson equation
\begin{align}
\label{PG}
a^2\Delta G=G^1\,.
\end{align}
Substituting Eq.~(\ref{G-deco}) into Eq.~(\ref{LG}) and using Eq.~(\ref{G-bi}),
we obtain for the gradient part $G^1$ the following equation
\begin{align}
\label{LG1}
L G^1=a^2 \Delta G^0=a^2 \delta(\Bx-\Bx')\,.
\end{align}
Therefore, the regularized Green function $G$, which is the Green function of
the Helmholtz-Laplace operator, can be represented as a superposition of the Green
functions of the Laplace and Helmholtz operators.
On the other hand, it follows from Eqs.~(\ref{LG})--(\ref{LG1}) that 
$G$ can be written as the convolution of the Green functions 
of the Laplace and Helmholtz operators into the following way
\begin{align}
G=\frac{1}{a^2}\, G^1*G^0\,,
\end{align}
satisfying
\begin{align}
L\Delta G=\frac{1}{a^2}\,L\Delta(G^1*G^0)
=\frac{1}{a^2}\,(L G^1)*(\Delta G^0)=
\delta*(\Delta G^0)=\delta(\Bx-\Bx')\,,
\end{align}
where Eqs.~(\ref{G0-pde}) and (\ref{LG1}) have been used.

Thus, $A_k^0$, $H_k^0$, $B_k^0$ and $G^0$ are the classical fields (or Maxwell fields)
and the fields $A_k^1$, $H_k^1$, $B_k^1$ and $G^1$ are the gradients parts
which are the non-classical fields (or static Proca or Yukawa fields) 
depending on the parameter $a$.

\subsection{An electric current loop and the Biot-Savart law}
We consider a closed electric circuit $C$ carrying the steady current $I$. 
The electric current vector density (``true current'') of such a closed loop
is given by
\begin{align}
\label{J-L}
J_{k}&=I\, \delta_k(C)=I \oint_C \delta(\Bx-\Bx')\, d l'_k\,.
\end{align}
Here, $\delta_j(C)$ is the Dirac delta function for a closed curve $C$.
Substituting Eq.~(\ref{J-L}) into Eq.~(\ref{A-int}), the magnetic vector gauge potential
of a closed loop is 
\begin{align}
\label{A-L-L}
A_k=\frac{\mu_0 I}{4\pi}\,\oint_C\frac{1}{R} \Big(1-\e^{-R/a}\Big)\, d l'_k\,.
\end{align}
It is important to note that the field~(\ref{A-L-L}) is non-singular. 
According to the ``Bifield'' ansatz~(\ref{A-deco}) the magnetic vector gauge potential~(\ref{A-L-L})
may be decomposed into the classical part
\begin{align}
\label{A-L-0}
A_k^0=\frac{\mu_0 I}{4\pi}\,\oint_C\frac{1}{R} \, d l'_k
\end{align}
and the gradient part 
\begin{align}
\label{A-L-1}
A_k^1=-\frac{\mu_0 I}{4\pi}\,\oint_C\frac{\e^{-R/a}}{R}\, d l'_k\,.
\end{align}
Both $A_k^0$ and $A_k^1$ are singular.
In general, the field $A_k=A_k^0+A_k^1$ which is the sum 
of a long-ranging ``Coulomb-like'' field $A_k^0$ and 
a short-ranging ``Yukawa-like'' field $A_k^1$ is non-singular.

For a closed electric current loop,
the Biot-Savart law valid in gradient magnetostatics is calculated as
\begin{align}
\label{B-BS}
B_i&=\frac{\mu_0 I}{4\pi}\,\epsilon_{ijk}
\pd_j \oint_C\frac{1}{R} \Big(1-\e^{-R/a}\Big)\, d l'_k\nonumber\\
&=-\frac{\mu_0 I}{4\pi}\,\oint_C\epsilon_{ijk}\,\frac{R_j}{R^3} 
\bigg[1-\bigg(1+\frac{R}{a}\bigg)\e^{-R/a}\bigg]\, d l'_k\,.
\end{align}
Note that this field is finite in the whole space.
Eq.~(\ref{B-BS}) represents the magnetic field vector of a current loop
valid in the Bopp-Podolsky theory.
In the limit $a\rightarrow 0$, Eqs.~(\ref{A-L-L}) and (\ref{B-BS})
reduce to the ``classical'' results of magnetostatics (see, e.g., \citep{Jackson,Lorrain,PP}).
According to the ``Bifield'' ansatz~(\ref{B-deco}) the magnetic field vector~(\ref{B-BS})
might be decomposed into the classical part
\begin{align}
\label{B-L-0}
B_i^0=-\frac{\mu_0 I}{4\pi}\,\oint_C\epsilon_{ijk}\,\frac{R_j}{R^3}\,d l'_k\,
\end{align}
and the gradient part 
\begin{align}
\label{B-L-1}
B_i^1=\frac{\mu_0 I}{4\pi}\,\oint_C
\epsilon_{ijk}\,\frac{R_j}{R^3}\bigg(1+\frac{R}{a}\bigg)\e^{-R/a}\, d l'_k\,.
\end{align}
Both the long-ranging field $B_i^0$ and 
the short-ranging field $B_i^1$ are singular. However, the superposition
$B_i=B_i^0+B_i^1$ is non-singular.

From the point of view of the magnetic field~(\ref{B-BS}), the 
electric current is not anymore a $\delta$-string; the real electric current 
with ``core spreading'' is obtained by inserting Eq.~(\ref{J-L}) into Eq.~(\ref{J-true}) 
\begin{align}
\label{J-L-true}
J'_{k}&=\frac{\mu_0\, I}{4\pi a^2}\, \oint_C \frac{\e^{-R/a}}{R}\, d l'_k\,.
\end{align}
The Bopp-Podolsky length $a$ has the meaning of the region in which
non-local interaction is of fundamental importance.

The Lorentz force 
between an electric current $\BJ^{(A)}$ and a
magnetic field $\BB^{(B)}$ is given by (e.g.,~\citep{A97,AM})
\begin{align}
\label{P-L}
F^{(AB)}_m=\int_V \epsilon_{mli}\, B^{(B)}_i J^{(A)}_l\, d V\,.
\end{align}
If we substitute the electric current~(\ref{J-L}) of a loop $C^{(A)}$ and the magnetic
field~(\ref{B-BS}) of a loop $C^{(B)}$,
we obtain the interaction force between two loops $C^{(A)}$ and $C^{(B)}$ 
\begin{align}
\label{P-L2}
F^{(AB)}_m=-\frac{\mu_0\, I^{(A)} I^{(B)}}{4\pi}\,\oint_{C^{(A)}}\oint_{C^{(B)}}
\epsilon_{mli}
\epsilon_{ijk}\,\frac{R_j}{R^3} 
\bigg[1-\bigg(1+\frac{R}{a}\bigg)\e^{-R/a}\bigg]\, 
d l^{(B)}_k\, d l^{(A)}_l\,,
\end{align}
where $\BR=\Bx^{(A)}-\Bx^{(B)}$.
Eq.~(\ref{P-L2}) can be simplified and the force on a loop $C^{(A)}$ 
exerted by a loop $C^{(B)}$ is 
\begin{align}
\label{P-L3}
F^{(AB)}_j=-\frac{\mu_0\, I^{(A)} I^{(B)}}{4\pi}\,\oint_{C^{(A)}}\oint_{C^{(B)}}
\frac{R_j}{R^3} 
\bigg[1-\bigg(1+\frac{R}{a}\bigg)\e^{-R/a}\bigg]\, 
d l^{(B)}_i\, d l^{(A)}_i\,.
\end{align}
It follows that $F^{(AB)}_j=-F^{(BA)}_j$. 
Thus, it can be seen that the interaction force between two current loops is
non-singular in gradient magnetostatics.

Using the identity (see also~\citep{Gaete})
\begin{align}
\label{W-id}
\int_V \Big(B_i H_i+a^2\pd_k B_i\pd_k H_i\Big)\, d V
&=\int_V B_i \, L H_i \, d V +\text{div-term}\nonumber\\
&=\int_V (\epsilon_{ijk}\pd_j A_k) \, L H_i\, d V\nonumber\\
&=\int_V A_k\, (\epsilon_{kji} \pd_j L H_i)\, d V\nonumber\\
&=\int_V A_k J_k\, d V\,,
\end{align}
where we have used that the surface term vanishes at infinity, Eq.~(\ref{B-A}), 
partial integration, and the field equation~(\ref{FE-H}), we 
finally obtain the formula for the 
interaction energy between a current $\BJ^{(A)}$ and 
the magnetic vector gauge potential $\BA^{(B)}$:
\begin{align}
\label{W-AB}
W^{(AB)}=\int_V A^{(B)}_k J^{(A)}_k\, d V\,.
\end{align}
If we substitute the electric current~(\ref{J-L}) of a loop $C^{(A)}$ and the magnetic
vector gauge potential~(\ref{A-L-L}) of a loop $C^{(B)}$,
we find for the interaction energy between two loops $C^{(A)}$ and $C^{(B)}$ 
\begin{align}
\label{W-AB2}
W^{(AB)}&=I^{(A)}\oint_{C^{(A)}} A^{(B)}_k \, d l_k^{(A)}\nonumber\\
&=\frac{\mu_0\, I^{(A)} I^{(B)}}{4\pi}\,\oint_{C^{(A)}}\oint_{C^{(B)}}
\frac{1}{R}\,\Big(1-\e^{-R/a}\Big)\, d l^{(B)}_k\, d l^{(A)}_k\,.
\end{align}
This is the non-singular mutual interaction energy between two current loops.
In the limit $a\rightarrow 0$, Eq.~(\ref{W-AB2}) 
reduces to the ``classical'' singular result of magnetostatics 
(see, e.g., \citep{Jackson,Lorrain}).
If we define the mutual inductance between the loops $C^{(A)}$ and $C^{(B)}$ 
in gradient magnetostatics by
\begin{align}
\label{M-AB}
M^{(AB)}=\frac{\mu_0}{4\pi}\,\oint_{C^{(A)}}\oint_{C^{(B)}}
\frac{1}{R}\,\Big(1-\e^{-R/a}\Big)\, d l^{(B)}_k\, d l^{(A)}_k\,,
\end{align}
the interaction energy~(\ref{W-AB2}) can be written as 
\begin{align}
\label{W-AB3}
W^{(AB)}=I^{(A)}I^{(B)} M^{(AB)}\,.
\end{align}
It follows that $M^{(AB)}=M^{(BA)}$. Eq.~(\ref{M-AB}) is a purely geometric
quantity,  
which is the Neumann equation valid in gradient magnetostatics.
The self-energy of an electric current loop can be found by using the same
curve for $C^{(A)}$ and $C^{(B)}$, and inserting a factor $\frac{1}{2}$, so that,
$W^{(AA)}=\frac{1}{2}\, I^{(A)}I^{(A)} M^{(AA)}$, where $M^{(AA)}$ is the self-inductance.

\section{Gradient elasticity of Helmholtz type}

In this section, we investigate the theory of gradient elasticity of Helmholtz type.
The strain energy density of gradient elasticity theory  of Helmholtz type
for an isotropic, linearly elastic material has the form~\citep{LM05,GM10a,Lazar12,Lazar13}
\begin{align}
\label{W}
W=\frac{1}{2}\, C_{ijkl}\beta_{ij}\beta_{kl}
+\frac{1}{2}\, \ell^2 C_{ijkl}\pd_m \beta_{ij} \pd_m \beta_{kl}\, ,
%%%%\qquad\ell>0
\end{align}
where the tensor of the elastic moduli $C_{ijkl}$ is given by
\begin{align}
\label{C}
C_{ijkl}=\mu\big(
\delta_{ik}\delta_{jl}+\delta_{il}\delta_{jk}\big)
+\lambda\, \delta_{ij}\delta_{kl}\,.
\end{align}
Here, $\mu$ and $\lambda$ are the Lam\'e moduli and
$\beta_{ij}$ denotes the elastic distortion tensor\footnote{
Due to an existing confusion in the literature,
it is noted that $\beta_{ij}$  is the elastic distortion tensor of gradient elasticity 
and it should not be confused with the elastic distortion tensor $\beta_{ij}^0$ 
of classical elasticity.
}.
If the elastic distortion tensor is incompatible, it can be decomposed
as follows
\begin{align}
\label{beta}
\beta_{ij}=\pd_j u_i -\beta^\TP_{ij}\,,
\end{align}
where $u_i$ and $\beta^\TP_{ij}$ denote the displacement vector and the plastic
distortion tensor, respectively. In addition, $\ell$ is the material length scale 
parameter of gradient elasticity of Helmholtz type. 
For dislocations, $\ell$ is related to the size of the dislocation core.
The condition for non-negative strain energy density, $W\ge 0$, gives
for the material moduli the following relations
\begin{align}
\label{cond-l}
(2\mu+3\lambda)\ge 0\,, \qquad\mu\ge 0\,,\qquad\ell^2\ge 0\,.
\end{align}

Defects like dislocations may be the
reason that the elastic and plastic distortion tensors are incompatible.
Since dislocations cause self-stresses, body forces are zero.
The dislocation density tensor can be defined in terms of the elastic and plastic 
distortion tensors as follows (e.g.,~\citep{Kroener58})
\begin{align}
\label{DD-el}
\alpha_{ij}&=\epsilon_{jkl}\pd_k \beta_{il}\,,\\
\label{DD-pl}
\alpha_{ij}&=-\epsilon_{jkl}\pd_k \beta^\TP_{il}\,,
\end{align}
which fulfills the following Bianchi identity 
\begin{align}
\label{BI}
\pd_j \alpha_{ij}=0\,.
\end{align}
It means that dislocations do not end inside the body. 
From Eq.~(\ref{DD-pl}) it can be seen that the plastic distortion tensor,
which plays the role of eigendeformation and eigenstrain,
cannot be neglected for dislocations.

From Eq.~(\ref{W}) it follows that the corresponding constitutive relations are
\begin{align}
\label{CR1}
\sigma_{ij}&=\frac{\pd W}{\pd \beta_{ij}}
=\frac{\pd W}{\pd e_{ij}}
=C_{ijkl}\beta_{kl}=C_{ijkl}e_{kl}\,,\\
\label{CR2}
\tau_{ijk}&=\frac{\pd W}{\pd (\pd_k\beta_{ij})}
=\frac{\pd W}{\pd (\pd_k e_{ij})}
=\ell^2\,C_{ijmn}\pd_k \beta_{mn}=\ell^2\pd_k \sigma_{ij}\,.
\end{align}
Here, $\sigma_{ij}=\sigma_{ji}$ is the Cauchy stress tensor\footnote{
In order to avoid the existing confusion and non-unique terminology 
in the literature of gradient elasticity
(e.g.,~\citep{Polyzos,Polyzos2,Karlis,Aifantis11}),
it has to be noted that $\sigma_{ij}$  and $e_{ij}$ are the Cauchy stress
tensor and the elastic strain tensor of gradient elasticity 
and they should not be confused with the Cauchy stress tensor 
$\sigma_{ij}^0$  and the elastic strain tensor $e_{ij}^0$ 
of classical elasticity. 
On the other hand, \citet{Georg04} used the notation
of monopolar stress tensor
for $\sigma_{ij}$ and dipolar stress tensor for $\tau_{ijk}$.
\citet{Georg06} used the terminology: 
$\sigma_{ij}$ is the monopolar (or Cauchy in the nomenclature 
of Mindlin~\citep{Mindlin64}) stress tensor and $\tau_{ijk}$ is the
dipolar (or double) stress tensor.
},
$\tau_{ijk}=\tau_{jik}$ is the so-called double stress tensor, 
and $e_{ij}=1/2(\beta_{ij}+\beta_{ji})$ is the elastic strain tensor
(see also~\citep{Eshel,Eshel73,LM05,Lazar13,GM10b,Georg00}).
Using Eqs.~(\ref{CR1}) and (\ref{CR2}), Eq.~(\ref{W}) 
can also be written as~\citep{LM05} 
\begin{align}
\label{W3}
W=\frac{1}{2}\, \sigma_{ij}e_{ij}
+\frac{1}{2}\, \ell^2 \pd_k \sigma_{ij} \pd_k e_{ij}\, .
%%%%\qquad\ell>0
\end{align}
It is obvious that the strain energy density~(\ref{W3}) 
exhibits a  ``stress-strain'' symmetry both in
$\sigma_{ij}$ and $e_{ij}$ and in $\pd_k \sigma_{ij}$ and $\pd_k e_{ij}$.
In addition, it yields
\begin{align}
\frac{\pd^2 W}{\pd e_{ij}\, \pd e_{kl}}=C_{ijkl}\,,\qquad\quad
\frac{\pd^2 W}{\pd (\pd_m e_{ij})\, \pd (\pd_m e_{kl})}=\ell^2 C_{ijkl}\,.
\end{align}

Using a variational principle, the Euler-Lagrange equation 
reads for gradient elasticity (see, e.g.,~\citep{Maugin93,KA02,LA07,AL09})
\begin{align}
\label{EL-u}
\frac{\delta W}{\delta u_i}=\frac{\pd W}{\pd u_i}
-\pd_j\, \frac{\pd W}{\pd (\pd_j u_i)}
+\pd_k \pd_j\, \frac{\pd W}{\pd (\pd_k \pd_j u_i)}=0\,.
\end{align}
For vanishing body forces and using the constitutive
relations~(\ref{CR1}) and (\ref{CR2}),
the Euler-Lagrange equation~(\ref{EL-u}) takes the following form 
in terms of the Cauchy and double stress tensors~(e.g.,~\citep{Mindlin64})
\begin{align}
\label{EC-0}
\pd_j (\sigma_{ij}-\pd_k \tau_{ijk})=0\,.
\end{align}
Using the relation that the double stress tensor is the gradient 
of the Cauchy stress tensor in Eq.~(\ref{CR2}), then Eq.~(\ref{EC-0}) reduces to
(e.g.,~\citep{Lazar13})
\begin{align}
\label{EC}
L \pd_j \sigma_{ij}=0\, ,
\end{align}
where now 
\begin{align}
\label{L}
L=1-\ell^2\Delta 
\end{align}
is the Helmholtz operator, depending on the gradient length scale $\ell$. 
It is interesting to note that the equilibrium condition~(\ref{EC}) is similar 
in the form to the generalized Coulomb gauge condition~(\ref{CG1}).

If we substitute the constitutive relation~(\ref{CR1}) and 
Eq.~(\ref{beta}) into the equilibrium condition~(\ref{EC}), 
we obtain the following inhomogeneous Helmholtz-Navier equation for the
displacement vector $\Bu$
\begin{align}
\label{u-L0}
L L_{ik} u_k =C_{ijkl} \pd_j L  \beta^{\TP}_{kl}\,,
\end{align}
where $L_{ik}=C_{ijkl}\pd_j\pd_l$ is the differential operator of the Navier 
equation.  
For an isotropic material, it reads
\begin{align}
L_{ik}=\mu\, \delta_{ik}\Delta+(\mu+\lambda)\, \pd_i\pd_k\,.
\end{align}
Eq.~(\ref{u-L0}) is nothing but the equilibrium condition~(\ref{EC})
written in terms of the displacement vector $\Bu$ and the plastic distortion tensor 
$\Bbeta^\TP$.
From Eq.~(\ref{u-L0})
we can also derive an inhomogeneous Helmholtz-Navier equation for 
the elastic distortion tensor $\Bbeta$
\begin{align}
\label{B-L0}
L L_{ik}\beta_{km} =-C_{ijkl}\epsilon_{mlr} \pd_j L \alpha_{kr}\,,
\end{align}
where the dislocation density tensor $\Balpha$ is the source field.
It is interesting to note that Eqs.~(\ref{u-L0}) and (\ref{B-L0}) 
have a similar form as Eq.~(\ref{FE-H-true}).

On the other hand, adopting the so-called 
``Ru-Aifantis theorem''~\citep{RA} in terms of stresses, 
Eq.~(\ref{EC}) can be written as an equivalent system of two equations, namely
\begin{align}
\label{RA1}
&\pd_j \sigma_{ij}^0=0\,,\\
\label{RA2}
& L \sigma_{ij}=\sigma_{ij}^0\,,
\end{align}
where $\sigma^0_{ij}$ is the classical Cauchy stress tensor (sometimes also called 
``total stress'' tensor~\citep{VS95,Vardu,Ex,Polyzos} 
or the ``polarization'' of the stress $\sigma_{ij}$~\citep{Jaunzemis}).
Although, \citet{Vardu} and \citet{Ex} called the stress $\sigma_{ij}^0$ as total
stress tensor in the framework of gradient elasticity, 
their obtained mode-III and mode-I crack solutions for $\sigma_{ij}^0$ do not depend
on the gradient parameter~$\ell$.
In fact, using gradient elasticity,
the solution for the stress $\sigma_{ij}^0$ of a mode-III crack~\citep{Vardu} 
agrees with the mode-III crack solution given by~\citet{AA92,AA97} in the framework of
gradient elasticity.
However, as it is already mentioned by~\citet{AA92,AA97},
the solution of the stress of the mode-III crack is the same as the stress field 
of a mode-III crack in the classical theory of elasticity and 
it is singular at the crack tip. 
In addition, in  a formal sense, Eqs.~(\ref{RA1}) and (\ref{RA2})
are  similar to Eqs.~(\ref{CG0}) and (\ref{LA}), respectively.
Therefore, the tensor $\sigma_{ij}^0$ should be identified with the classical 
stress tensor. 

As shown by~\citet{LM05,LM06}, the following Helmholtz equations (pdes of 2nd-order)
for the elastic distortion tensor, the displacement vector, the plastic
distortion tensor, and the dislocation density tensor
can be derived from the inhomogeneous Helmholtz equation~(\ref{RA2}) 
\begin{align}
\label{B-H}
&L \beta_{ij} =\beta_{ij}^0\,,\\
\label{u-H}
&L u_i =u_i^0\,,\\
\label{BP-H}
&L \beta^\TP_{ij} =\beta_{ij}^{\TP,0}\,,\\
\label{A-H}
&L \alpha_{ij} =\alpha_{ij}^0\,,
\end{align}
where $\Bbeta^0$, $\Bu^0$, $\Bbeta^{\TP,0}$ and $\Balpha^0$ are the
corresponding classical fields.
Note that the fields $\Bbeta^0$, $\Bu^0$, $\Bbeta^{\TP,0}$ and $\Balpha^0$ are 
singular and they are the sources in the inhomogeneous Helmholtz equations~(\ref{B-H})--(\ref{A-H}).
Using the Helmholtz equations (\ref{BP-H}) and (\ref{A-H}),
the Helmholtz-Navier equations~(\ref{u-L0}) and (\ref{B-L0}) can be simplified to
the following inhomogeneous Helmholtz-Navier equations (pdes of 4th-order)
\begin{align}
\label{u-L}
&L L_{ik} u_k =C_{ijkl}\pd_j \beta^{\TP,0}_{kl}\,,\\
\label{B-L}
&L L_{ik}\beta_{km} =-C_{ijkl}\epsilon_{mlr}\pd_j \alpha_{kr}^0\,,
\end{align}
where now the classical plastic distortion tensor $\Bbeta^{\TP,0}$ and the 
classical dislocation density tensor $\Balpha^0$ are the source fields for the 
displacement vector $\Bu$ and the elastic distortion tensor $\Bbeta$, respectively.
The important type of pde for dislocations in gradient elasticity 
is the Helmholtz-Navier equation, which is a pde of 4th-order.
Using the technique of Green functions (e.g.,~\citep{Barton,Barut}), 
Eqs.~(\ref{u-L}) and (\ref{B-L})
can be easily solved for any given sources  $\Bbeta^{\TP,0}$ and $\Balpha^0$.
This can be considered as an eigenstrain problem of dislocations in the
framework of gradient elasticity.

\subsection{Green tensor of the three-dimensional Helmholtz-Navier equation} 

The Green tensor of the three-dimensional Helmholtz-Navier equation 
is defined by
\begin{align}
\label{pde-HN}
&L L_{ik} G_{kj} =-\delta_{ij} \delta(\Bx-\Bx')
\end{align}
and is given by~\citep{Lazar13}
\begin{align}
\label{G}
G_{ij}(R)=\frac{1}{16\pi\mu(1-\nu)}\, \Big[2(1-\nu)\delta_{ij}\Delta-
\pd_i\pd_j\Big] A(R)\,,
\end{align}
with the ``regularization function''
\begin{align}
\label{A}
A(R)=R+\frac{2\ell^2}{R}\,
\Big(1-\e^{-R/\ell}\Big)\,, 
\end{align} 
where $R = |\Bx-\Bx'|$ and $\nu$ is the Poisson ratio.
In the limit $\ell\rightarrow 0$, the three-dimensional Green tensor
of classical elasticity~\citep{Mura,Li} 
is recovered from Eqs.~(\ref{G}) and (\ref{A}). 
In contrast to the Green tensor of the Navier equation, which is singular, the
Green tensor of the Helmholtz-Navier equation is non-singular (see also~\citep{Lazar13}).
Thus, Eq.~(\ref{G}) represents the regularized Green tensor in the gradient elasticity theory
of Helmholtz type.
It is noted that $A(R)$ can be written as the convolution of $R$ and $G(R)$
\begin{align}
A(R)=R*G(R)\,,
\end{align}
where $G$ is here the Green function of the three-dimensional Helmholtz equation
\begin{align}
\label{pde-H}
L G =\delta(\Bx-\Bx')\,,
\end{align}
which is given by
\begin{align}
\label{G-H}
G(R)=\frac{1}{4\pi \ell^2 R}\, \e^{-R/\ell}\,.
\end{align}
The Green function~(\ref{G-H}) is a Dirac-delta sequence 
with parametric dependence $\ell$
\begin{align}
\lim_{\ell\to 0} G(R)=\delta(\BR)
\end{align}
and it plays the role of the ``regularization Green function'' 
in gradient elasticity. 
In fact, $G(R)$ gives an isotropic regularization 
in the theory of isotropic gradient elasticity.

In addition, it holds
\begin{align}
\Delta\Delta\, R=-8\pi\, \delta(\Bx-\Bx')\,.
\end{align}
The ``regularization function''~(\ref{A}) fulfills the relations
\begin{align}
\label{A-LDD}
L \Delta\Delta A(R)&=-8\pi\, \delta(\Bx-\Bx')\,,\\
\label{A-DD}
\Delta\Delta A(R)&=-8\pi\, G(R)\,,\\
\label{A-L}
L A(R)&=R\,.
\end{align}
Thus, $A(R)$ is the Green function of Eq.~(\ref{A-LDD})
which is a three-dimensional Helmholtz-bi-Laplace equation (pde of 6th-order).

In addition, the Green tensor~(\ref{G}) satisfies the following relation
\begin{align}
\label{LG-H}
L G_{ij}(R)&=\frac{1}{16\pi\mu(1-\nu)}\, \Big[2(1-\nu)\delta_{ij}\Delta-
\pd_i\pd_j\Big] L A(R)\nonumber\\
&=\frac{1}{16\pi\mu(1-\nu)}\, \Big[2(1-\nu)\delta_{ij}\Delta-\pd_i\pd_j\Big] R\nonumber\\
&=G_{ij}^0(R)\,,
\end{align}
where $G_{ij}^0$ is the 
Green tensor of the ``classical'' Navier equation,
$L_{ik} G^0_{kj} =-\delta_{ij} \delta(\Bx-\Bx')$.
Eq.~(\ref{LG-H}) is an inhomogeneous Helmholtz equation. 
A consequence of  Eq.~(\ref{LG-H}) is that the Green tensor of 
the Helmholtz-Navier equation may be written as the convolution of the Green
function of the Helmholtz equation with the ``classical'' Green tensor 
of the Navier equation
\begin{align}
\label{GT-conv}
G_{ij}=G*G_{ij}^0\,.
\end{align}
Therefore, the Green tensor, $G_{ij}$, fulfills the following inhomogeneous pdes
\begin{align}
\label{LG-N}
L_{ik} G_{kj}& =L_{ik}(G*G_{kj}^0)= G*(L_{ik} G_{kj}^0)=-\delta_{ij}\, G\,,\\
\label{LG-L}
L G_{ij} &=L(G*G_{ij}^0)= G^0_{ij}*(L G)=G^0_{ij}\,.
\end{align}
Eq.~(\ref{LG-N}) is an inhomogeneous Navier equation and 
Eq.~(\ref{LG-L}) is an inhomogeneous Helmholtz equation 
for the Green tensor of gradient elasticity of Helmholtz type.
In addition, using the convolution representation~(\ref{GT-conv}) and 
Eq.~(\ref{LG-N}), it can be easily seen that Eq.~(\ref{pde-HN})
is satisfied
\begin{align}
\label{LLG-}
L L_{ik} G_{kj}& =LL_{ik}(G*G_{kj}^0)= (LG)*(L_{ik} G_{kj}^0)=-\delta_{ij}\, 
LG=-\delta_{ij}\delta(\Bx-\Bx')\,.
\end{align}

\subsection{``Ru-Aifantis theorem'' for dislocations and the Bifield-Ansatz} 

Originally, the so-called ``Ru-Aifantis theorem''~\citep{RA}
was derived for compatible gradient elasticity. 
The ``Ru-Aifantis theorem'' may be used for problems 
concerning bodies of infinite extent.  
In fact, the so-called ``Ru-Aifantis theorem'' is a special case of a more
general technique; well-known in the theory of partial differential equations  
(see, e.g.,~\citep{Vekua}).
Such an approach is mainly based on the decomposition 
of a pde of higher-order into a system of pdes
of lower-order and on the property that the appearing differential 
operator(s) can be written as a product of differential operators of lower-order
(operator-split).
Also, the property that the differential operators commute is often used in the 
operator-split.
Here, we give the generalization of
such a technique towards the (incompatible) theory of dislocations in gradient elasticity. 
The difficulty in the theory of dislocations in the framework of gradient elasticity
is that both fields, the field on the left hand side and the source field 
on the right hand side of the Helmholtz-Navier equations~(\ref{u-L0}) and
(\ref{B-L0}) are ``a priori'' unknown. 
Therefore, the ``Ru-Aifantis theorem'' valid for one single field has
to be generalized towards two fields. For that reason the number of equations
of the system is changed from two to three. 
However, it is possible to obtain also equivalent versions of a system with
only two equations.
We give here both equivalent versions.
It is noted that all the equations derived in this subsection 
are valid for isotropic as well as
anisotropic gradient elasticity of Helmholtz type.

The inhomogeneous Helmholtz-Navier equation~(\ref{u-L0}) 
for the displacement field with 
the plastic distortion tensor as source (pde of 4th-order)
\begin{align}
\label{u-L0-RA}
L L_{ik} u_k =C_{ijkl} \pd_j L \beta^{\TP}_{kl}\,
\end{align}
can be decomposed into the following system of partial differential equations;
namely into an inhomogeneous Navier equation (pde of 2nd-order)
\begin{align}
\label{u-L0-RA0}
L_{ik} u_k^0 =C_{ijkl} \pd_j  \beta^{\TP,0}_{kl}\,
\end{align}
and into two uncoupled inhomogeneous Helmholtz equations (pdes of 2nd-order)
\begin{align}
\label{u-H-RA}
&L u_i =u_i^0\,,\\
\label{BP-H-RA}
&L \beta^\TP_{ij} =\beta_{ij}^{\TP,0}\,.
\end{align}
Eq.~(\ref{u-L0-RA0}) 
is the classical Navier equation known from dislocation theory,
which serves the source fields for Eqs.~(\ref{u-H-RA}) and (\ref{BP-H-RA}).
If we substitute Eq.~(\ref{u-H-RA}) into Eq.~(\ref{u-L0-RA0}), 
we recover the Helmholtz-Navier equation~(\ref{u-L}).
Substituting Eqs.~(\ref{u-H-RA}) and (\ref{BP-H-RA}) into
Eq.~(\ref{u-L0-RA0}), Eq.~(\ref{u-L0-RA}) is recovered.

In addition, Eq.~(\ref{u-L0-RA}) may be rewritten equivalently 
into the following system of pdes
\begin{align}
\label{u-L0-RA0-1}
L L_{ik} u_k &=C_{ijkl} \pd_j  \beta^{\TP,0}_{kl}\,,\\
\label{B-H-RA0-1}
L \beta^\TP_{ij} &=\beta_{ij}^{\TP,0}\,,
\end{align}
or into the system of pdes
\begin{align}
\label{u-L0-RA0-2}
 L_{ik} u^0_k &=C_{ijkl} \pd_j  L \beta^{\TP}_{kl}\,,\\
\label{u-H-RA0-2}
L u_i &=u_i^0\,.
\end{align}

On the other hand,
the inhomogeneous Helmholtz-Navier equation~(\ref{B-L0}) 
for the elastic distortion tensor with 
the dislocation density tensor as source (pde of 4th-order)
\begin{align}
\label{B-L0-RA}
L L_{ik}\beta_{km} =-C_{ijkl}\epsilon_{mlr} \pd_j L \alpha_{kr}\,,
\end{align}
can be decomposed into the following system of partial differential equations;
namely into an inhomogeneous Navier equation (pde of 2nd-order)
\begin{align}
\label{B-L0-RA0}
L_{ik}\beta_{km}^0 =-C_{ijkl}\epsilon_{mlr} \pd_j \alpha^0_{kr}\,,
\end{align}
and into two inhomogeneous Helmholtz equations (pdes of 2nd-order)
\begin{align}
\label{B-H-RA}
&L \beta_{ij} =\beta_{ij}^0\,,\\
\label{A-H-RA}
&L \alpha_{ij} =\alpha_{ij}^0\,.
\end{align}
Eq.~(\ref{B-L0-RA0}) 
is a classical Navier equation known from dislocation theory,
which serves the source fields for Eqs.~(\ref{B-H-RA}) and (\ref{A-H-RA}).
If we substitute Eq.~(\ref{B-H-RA}) into Eq.~(\ref{B-L0-RA0}), 
we recover the Helmholtz-Navier equation~(\ref{B-L}).

In addition, Eq.~(\ref{B-L0-RA}) can be rewritten equivalently 
into the following system of pdes
\begin{align}
\label{B-L0-RA0-1}
L L_{ik}\beta_{km} &=-C_{ijkl}\epsilon_{mlr} \pd_j \alpha^0_{kr}\,,\\
\label{A-H-RA-1}
L \alpha_{ij} &=\alpha_{ij}^0\,,
\end{align}
or into the system of pdes
\begin{align}
\label{B-L0-RA0-2}
L_{ik}\beta_{km}^0 &=-C_{ijkl}\epsilon_{mlr} \pd_j L \alpha_{kr}\,,\\
\label{B-H-RA-2}
L \beta_{ij} &=\beta_{ij}^0\,.
\end{align}

Using the Ru-Aifantis approach for the stress tensor, the equilibrium condition (pde of 3rd-order)
\begin{align}
L \pd_j \sigma_{ij}=0
\end{align}
is decomposed into the following system of two equations (pdes of 1st-order and 2nd-order)
\begin{align}
\label{FE-Bi-0}
&\pd_j \sigma_{ij}^0=0\,,\\
\label{RA0}
&L \sigma_{ij}=\sigma_{ij}^0\,.
\end{align}
In (linear) gradient elasticity, 
the ``Bifield'' ansatz,
as it has been described in subsection~(2.1) for the
theory of gradient magnetostatics, 
reads for the stress tensor 
\begin{align}
\label{t-bi}
\sigma_{ij}=\sigma_{ij}^0+\sigma_{ij}^1\,.
\end{align}
Substituting Eq.~(\ref{t-bi}) into the Helmholtz equation~(\ref{RA0}),
the following Helmholtz equation for the gradient part
of the stress tensor $\sigma_{ij}^1$ is obtained
\begin{align}
\label{FE-Bi-1}
L \sigma_{ij}^1=\ell^2 \Delta \sigma_{ij}^0\,,
\end{align}
where the Laplacian of the classical stress tensor $\sigma_{ij}^0$ is the
inhomogeneous part.
Moreover, the following Poisson equation for $\sigma_{ij}$ can be obtained 
by inserting Eq.~(\ref{t-bi}) into the Helmholtz equation~(\ref{RA0})
\begin{align}
\label{PT}
\ell^2\Delta\sigma_{ij}=\sigma^1_{ij}\,.
\end{align}
Thus, $\sigma_{ij}^1$ is a kind of relative stress tensor 
which is equilibrated by the double stress tensor~(\ref{CR2}) (see also~\citep{VS95,Vardu})
\begin{align}
\sigma_{ij}^1=\pd_k \tau_{ijk}\,.
\end{align}
The ``Bifield'' ansatz of the Cauchy stress tensor~(\ref{t-bi}) 
induces a ``Bifield decomposition for the double stress tensor~(\ref{CR2})
\begin{align}
\label{tau-bi}
\tau_{ijk}=\tau_{ijk}^0+\tau_{ijk}^1\,
\end{align}
with (see also~\citep{Ara})
\begin{align}
\tau^0_{ijk}&=\ell^2 \pd_k \sigma^0_{ij}\,,\\
\tau^1_{ijk}&=\ell^2 \pd_k \sigma^1_{ij}\,.
\end{align}

The Ru-Aifantis approach for the elastic distortion tensor
decomposes the equilibrium condition (pde of 3rd-order)
\begin{align}
C_{ijkl} L \pd_j \beta_{kl}=0
\end{align}
into the following system of two equations (pdes of 1st-order and 2nd-order)
\begin{align}
&C_{ijkl} \pd_j \beta_{kl}^0=0\,,\\
\label{RA-B}
&L  \beta_{ij}=\beta_{ij}^0\,.
\end{align}
The ``Bifield'' ansatz for the elastic distortion tensor is given by
\begin{align}
\label{B-bi}
\beta_{ij}=\beta_{ij}^0+\beta_{ij}^1\,.
\end{align}
The substitution of Eq.~(\ref{B-bi}) into the Helmholtz equation~(\ref{RA-B})
gives the following Helmholtz equation for the gradient part
of the elastic distortion tensor $\beta_{ij}^1$
\begin{align}
L \beta_{ij}^1=\ell^2 \Delta \beta_{ij}^0\,,
\end{align}
where the Laplacian of the classical elastic distortion tensor $\beta_{ij}^0$ is
 the source term\footnote{\citet{Aifantis09,Aifantis09b} claimed that 
the gradient part, $e_{ij}^1$, of the elastic strain tensor of dislocations is determined 
from a homogeneous Helmholtz equation. This 
is obviously mistaken, since $e_{ij}^1$ satisfies the inhomogeneous Helmholtz equation:
$L e_{ij}^1=\ell^2 \Delta e_{ij}^0$, 
where $e_{ij}^0$ is the classical elastic strain tensor. 
}.
In addition,  if we substitute Eq.~(\ref{B-bi}) into the Helmholtz equation~(\ref{RA-B}), 
the following Poisson equation for $\beta_{ij}$ may be obtained 
\begin{align}
\label{PB2}
\ell^2\Delta\beta_{ij}=\beta^1_{ij}\,.
\end{align}

If we use a ``Bifield'' ansatz for the displacement vector
\begin{align}
u_{i}=u_{i}^0+u_{i}^1\,,
\end{align}
the inhomogeneous Helmholtz equation~(\ref{u-H-RA}) 
gives the following Helmholtz equation for the gradient part
of the displacement vector $u_{i}^1$
\begin{align}
L u_{i}^1 =\ell^2 \Delta u_{i}^0\,,
\end{align}
and the following Poisson equation for $u_i$
\begin{align}
\label{Pu}
\ell^2\Delta u_{i}=u^1_{i}\,.
\end{align}

For the ``regularization function''~(\ref{A}) 
the ``Bifield'' ansatz is 
\begin{align}
\label{A-bi}
A=A^0+A^1\,,
\end{align}
where 
\begin{align}
A^0=R\,,\qquad
A^1=\frac{2\ell^2}{R}\,
\Big(1-\e^{-R/\ell}\Big)\,.
\end{align}
In addition, the inhomogeneous Helmholtz equation~(\ref{A-L}) 
gives the following Helmholtz equation for the gradient part $A^1$
\begin{align}
L A^1 =\ell^2 \Delta A^0\,,
\end{align}
and the following Helmholtz-Laplace equation for the gradient part $A^1$
\begin{align}
L \Delta A^1 =- 8\pi \ell^2\, \delta(\BR)\,,
\end{align}
which shows that $A^1$ is the Green function of the Helmholtz-Laplace operator.
Moreover, $A$ satisfies the following Poisson equation
\begin{align}
\label{PA2}
\ell^2\Delta A=A^1\,.
\end{align}
Thus, using the ``Bifield'' ansatz, it can be seen that 
$\beta_{ij}^0$, $\sigma_{ij}^0$, $u_{i}^0$ and $A^0$ are the classical fields 
and $\beta_{ij}^1$, $\sigma_{ij}^1$ , $u_{i}^1$ and $A^1$
are the gradient parts depending on the gradient parameter~$\ell$.

An important consequence of this procedure 
is that the tensor $\sigma_{ij}^0$ is identified with the classical stress tensor
and that the tensor $\sigma_{ij}^1$, which is the gradient part of the stress, 
corresponds to the relative stress tensor. 
If the classical fields are known, 
the gradient parts are the only unknown fields in gradient theory.
Moreover, the gradient parts are given by 
inhomogeneous Helmholtz equations. 
The classical fields only satisfy the field equations of classical elasticity.
No Helmholtz equation where a Helmholtz operator $L$ acting on 
the classical fields is part of the theory 
of gradient elasticity\footnote{
Using an erroneous terminology in gradient elasticity, \citet{Polyzos,Karlis,Gianna}
and \citet{Aifantis11} derived
an inhomogeneous Helmholtz equation for the classical Cauchy stress tensor:
$L\sigma^0_{ij}=\sigma_{ij}$, which is based on a physical misinterpretation 
of the Cauchy stress tensor in gradient elasticity.
}.
In general, both the classical fields and the gradient parts can be singular,
only the superposition of the classical and the gradient parts gives
non-singular fields due to a ``physical'' regularization.
The physical interpretation of the fields in gradient elasticity of Helmholtz
type is in agreement with the physical interpretation of the fields 
in the Bopp-Bodolsky theory.

\subsection{Dislocation loops}
\label{DL}
In this subsection, we consider a dislocation loop in an unbounded body
in the framework of
gradient elasticity theory of Helmholtz type.
All the dislocation key-formulas are derived for gradient elasticity
of Helmholtz type.
For a general dislocation loop $C$, the 
classical dislocation density and the plastic distortion
tensors read~ (e.g.,~\citep{deWit2,Kossecka74})
\begin{align}
\label{A0}
\alpha^0_{ij}&=b_i\, \delta_j(C)=b_i \oint_C \delta(\Bx-\Bx')\, d l'_j\,,\\
%%\end{align}
%%\begin{align}
\label{B0}
\beta^{\TP, 0}_{ij}&=-b_i\, \delta_j(S)=-b_i \int_S \delta(\Bx-\Bx')\, d S'_j\,,
\end{align}
where $b_i$ is the Burgers vector,  $d l'_j$ denotes the dislocation line element at
$\Bx'$ and $d S'_j$ is the corresponding dislocation loop area.
The surface $S$ is the dislocation surface, which is a ``cap'' of the 
dislocation line $C$.
The surface $S$ represents the area swept by the loop $C$ during its
motion.
The plastic distortion~(\ref{B0}) caused by a dislocation loop is concentrated
at the dislocation surface $S$. 
Thus, the surface $S$ is what determines the history of the plastic distortion
of a dislocation loop.
$\delta_j(C)$ is the Dirac delta function for a closed curve $C$ and
$\delta_j(S)$ is the Dirac delta function for a surface $S$ 
with boundary $C$.

The solution of Eq.~(\ref{A-H}) is the following convolution
integral
\begin{align}
\label{A-grad0}
\alpha_{ij}=G*\alpha_{ij}^0
=b_i \oint_C G(R)\, d l'_j\,,
\end{align}
where $G(R)$ denotes
the three-dimensional Green function of the Helmholtz equation given 
by Eq.~(\ref{G-H}). 
The explicit solution of the dislocation density tensor for a dislocation loop
reads
\begin{align}
\label{A-grad}
\alpha_{ij}(\Bx)=\frac{b_i}{4\pi \ell^2 }\,\oint_C \frac{\e^{-R/\ell}}{R}\, d l'_j
\,,
\end{align}
describing a dislocation core spreading.
If we compare Eqs.~(\ref{A0}) and (\ref{A-grad}) with
Eqs.~(\ref{J-L}) and (\ref{J-L-true}), respectively,
it can be seen that $\Balpha^0$ plays the role of the ``true'' dislocation density tensor
and $\Balpha$ has the physical meaning of an ``effective'' dislocation
density tensor.

The plastic distortion tensor of a dislocation loop, which is the solution
of Eq.~(\ref{BP-H}), is given by the convolution integral
\begin{align}
\label{BP-grad0}
\beta^\TP_{ij}=G*\beta_{ij}^{\TP,0}
=-b_i \int_S G(R)\, d S'_j\, .
\end{align}
Explicitly, it reads 
\begin{align}
\label{BP-grad}
\beta^\TP_{ij}(\Bx)=-\frac{b_i}{4\pi \ell^2 }\,\int_S \frac{\e^{-R/\ell}}{R}\, d S'_j\, .
\end{align}
It is important to note that the gradient solution of the plastic
distortion is not concentrated at the dislocation surface $S$,
but it is distributed around $S$ according to Eq.~(\ref{BP-grad}).
The field $\Bbeta^\TP$ may be also called the ``effective'' plastic distortion.
Substituting Eq.~(\ref{BP-grad}) in Eq.~(\ref{DD-pl}) and using
the Stokes theorem, formula~(\ref{A-grad}) is recovered.
Due to the convolution of the classical dislocation density, $\Balpha^0$, and the
classical plastic distortion, $\Bbeta^{\TP,0}$, with the Green function, $G$,
the effective dislocation density, $\Balpha$, and the
effective plastic distortion, $\Bbeta^{\TP}$, are smeared out and 
modeling, in such a manner, a dislocation core region in gradient
elasticity.
In this way, the dislocation core spreading function, $G$, is of Yukawa type. 
For small distances, $R\ll\ell$, $G$ varies as $1/R$ and for larger distances,
however, $G$ decreases exponentially. Therefore, 
the dislocation core spreading function has a finite range.

\subsubsection{Burgers, Mura and Peach-Koehler stress formulas}

After a straightforward calculation
all the generalizations of the 
dislocation key-formulas (Mura, Peach-Koehler, and Burgers formulas)
towards gradient elasticity can be obtained.
Starting with the elastic distortion tensor of a dislocation loop,
the solution of Eq.~(\ref{B-L}) gives the representation as 
the following convolution integral
\begin{align} 
\label{B-Mura0}
\beta_{im}=\epsilon_{mnr} C_{jkln} \pd_k G_{ij}*\alpha^0_{lr}\,.
\end{align}
Eq.~(\ref{B-Mura0}) is the gradient version of ``Mura's half'' of the
so-called Mura-Willis formula~\citep{Mura63,Willis67}
due to the appearance of the Green tensor of the Helmholtz-Navier
equation~(\ref{G}).
Like in classical dislocation theory, 
the trace of the dislocation density
tensor $\alpha_{pp}^0$ gives zero contribution to the elastic 
distortion tensor, if we substitute $\alpha^0_{lr}=\delta_{lr}\,\alpha_{pp}^0$
into Eq.~(\ref{B-Mura0}).
Using the differentiation rule of the convolution~\citep{Wl,Kanwal}
and Eqs.~(\ref{A-H}) and (\ref{LG-H}), we find the identity
\begin{align}
\label{B-Mura0-id}
\beta_{im}=\epsilon_{mnr} C_{jkln} \pd_k G_{ij}*L \alpha_{lr}
=\epsilon_{mnr} C_{jkln} \pd_k L  G_{ij}*\alpha_{lr}
=\epsilon_{mnr} C_{jkln} \pd_k G^0_{ij}*\alpha_{lr}\,,
\end{align}
where $\beta^0_{im}=\epsilon_{mnr} C_{jkln} \pd_k G^0_{ij}*\alpha^0_{lr}$.
This shows again the relation between the four 
pdes~(\ref{B-L0-RA}), (\ref{B-L0-RA0}), (\ref{B-L0-RA0-1}), and (\ref{B-L0-RA0-2}).
Using Eq.~(\ref{A-grad0}), it can be represented as a double convolution
\begin{align}
\label{B-Mura0-2}
\beta_{im}=\epsilon_{mnr} C_{jkln} \pd_k G^0_{ij}*G*\alpha^0_{lr}
=\beta_{im}^0*G\,.
\end{align}
Finally, using the relation~(\ref{GT-conv}), 
Eq.~(\ref{B-Mura0}) is recovered from Eq.~(\ref{B-Mura0-2}).
Due to the Green function $G$, Eq.~(\ref{B-Mura0-2}) is the regularization of
the ``classical'' Mura equation.
If we substitute Eqs.~(\ref{C}) and (\ref{G}) into Eq.~(\ref{B-Mura0}), 
rearrange terms, use partial integration and $\pd_j\alpha_{ij}^0=0$,
we find for the elastic distortion tensor caused by the prescribed dislocation
distribution $\alpha_{kr}^0$
\begin{align}
\label{B-grad0}
\beta_{ij}(\Bx)&=-\frac{1}{8\pi}\int_V
\Big[\big(\epsilon_{jkl}\delta_{ir}-\epsilon_{rkl}\delta_{ij}
+\epsilon_{rij}\delta_{kl}\big)\pd_l \Delta
+\frac{1}{1-\nu} \,\epsilon_{rkl}\pd_l\pd_i\pd_j \Big] A(R)\, 
\alpha^0_{kr}(\Bx')\, d V'\, .
\end{align}
In the limit $\ell\rightarrow 0$, Eq.~(\ref{B-grad0}) tends
to the classical result given by~\citet{deWit3}.

Now, substituting the classical dislocation density tensor
of a dislocation loop (\ref{A0}) and carrying out the integration of
the delta function, we find the modified Mura formula
valid in gradient elasticity 
\begin{align}
\label{B-Mura}
\beta_{im}(\Bx)=\oint_C \epsilon_{mnr} b_l C_{jkln} \pd _k G_{ij}(R)\, d l'_r\,.
\end{align}
Substituting Eqs.~(\ref{C}) and (\ref{G}) into Eq.~(\ref{B-Mura}), 
rearranging terms and using the Stokes theorem or more directly from
Eq.~(\ref{B-grad0}), 
the generalized Mura equation valid in gradient elasticity is obtained 
\begin{align}
\label{B-grad}
\beta_{ij}(\Bx)&=-\frac{b_k}{8\pi}\oint_C
\Big[\big(\epsilon_{jkl}\delta_{ir}-\epsilon_{rkl}\delta_{ij}
+\epsilon_{rij}\delta_{kl}\big)\pd_l \Delta
+\frac{1}{1-\nu} \,\epsilon_{rkl}\pd_l\pd_i\pd_j \Big] A(R)\, 
d l'_r\, .
\end{align}
It is noted that 
if Eq.~(\ref{B-grad}) is substituted into Eq.~(\ref{DD-el}) and the 
relation~(\ref{A-DD}) is used, the dislocation density of a 
dislocation loop~(\ref{A-grad}) is recovered.
The symmetric part of the elastic distortion tensor (\ref{B-grad}) 
gives the elastic strain tensor of a dislocation loop
\begin{align}
\label{E-grad}
e_{ij}(\Bx)&=-\frac{b_k}{8\pi}\oint_C
\Big[\Big(
\frac{1}{2}\, \epsilon_{jkl}\delta_{ir}
+\frac{1}{2}\, \epsilon_{ikl}\delta_{jr}
-\epsilon_{rkl}\delta_{ij}\Big)\pd_l \Delta
+\frac{1}{1-\nu} \,\epsilon_{rkl}\pd_l\pd_i\pd_j \Big] A(R)\, 
d l'_r\, .
\end{align}

Using the constitutive relation~(\ref{CR1}) and Eq.~(\ref{B-Mura0-2}),
we obtain the representation of the Cauchy stress $\sigma_{ij}$ 
as convolution of the classical singular stress $\sigma_{ij}^0$ 
with the Green function $G$ of the Helmholtz equation
\begin{align}
\label{T-con}
\sigma_{ij}=\sigma_{ij}^0*G\,,
\end{align}
which is the (particular) solution of the inhomogeneous Helmholtz equation~(\ref{RA2}).
This follows by applying the Helmholtz operator~(\ref{L}) to both sides of Eq.~(\ref{T-con}).
The result is (see, e.g.,~\cite{Kanwal})
\begin{align}
\label{L-P}
L\sigma_{ij}=L(\sigma_{ij}^0*G)=\sigma_{ij}^0*(L G)=
\sigma_{ij}^0*\delta=\sigma_{ij}^0\,.
\end{align}
Such solution is unique in the class of generalized functions.
If we use Eq.~(\ref{T-con}), the property of the differentiation 
of a convolution and that the operation of convolution is 
commutative~\citep{Wl,Kanwal}, we find that the divergence of the 
Cauchy stress tensor in gradient elasticity is zero
\begin{align}
\label{DivT}
\pd_j \sigma_{ij}=\pd_j (G*\sigma_{ij}^0)=G*(\pd_j \sigma_{ij}^0)=0\,,
\end{align}
since $\pd_j \sigma_{ij}^0=0$. 
In order to differentiate a convolution, it suffices to differentiate any one of
the factors~\cite{Kanwal}.
Therefore, 
if the convolution~(\ref{T-con}) exists, then
the Cauchy stress tensor of gradient elasticity is self-equilibrated.
In addition, it can be seen that Eq.~(\ref{DivT}) is similar to the Coulomb 
gauge condition~(\ref{CG2}). 
Using the ``Bifield'' ansatz~(\ref{t-bi}), $\pd_j \sigma^1_{ij}=0$ follows
from Eq.~(\ref{DivT}).

Substituting Eq.~(\ref{E-grad}) into Eq.~(\ref{CR1})
and using Eq.~(\ref{C}), 
the non-singular stress field produced by a dislocation loop is 
\begin{align}
\label{T-grad}
\sigma_{ij}(\Bx)&=-\frac{\mu b_k}{8\pi}\oint_C
\Big[\big(\epsilon_{jkl}\delta_{ir}
+\epsilon_{ikl}\delta_{jr}\big)\pd_l \Delta
+\frac{2}{1-\nu}\, \epsilon_{rkl}\big(\pd_i\pd_j-\delta_{ij}\Delta\big)\pd_l
\Big] A(R)\, d l'_r\, ,
\end{align}
which can be interpreted as the Peach-Koehler stress formula within the framework of 
gradient elasticity.
One may verify that the Cauchy stress~(\ref{T-grad}) is 
divergence-less, $\pd_j\sigma_{ij}=0$, and thus
it is self-equilibrated.
The double stress tensor of a dislocation loop is easily obtained 
if Eq.~(\ref{T-grad}) is substituted into Eq.~(\ref{CR2}).
If we substitute Eqs.~(\ref{Aijk}) and (\ref{Aiik}) into
Eq.~(\ref{T-grad}), we  obtain the explicit expression for
the Peach-Koehler stress formula
\begin{align}
\label{T-grad2}
\sigma_{ij}(\Bx)&=-\frac{\mu b_l}{8\pi}\oint_C
\bigg[\Big(\epsilon_{jkl}\delta_{ir}
+\epsilon_{ikl}\delta_{jr}
-\frac{2}{1-\nu}\,\epsilon_{rkl}\delta_{ij}\Big)
\frac{2R_k}{R^3}\Big[1-\Big(1+\frac{R}{\ell}\Big)\e^{-R/\ell}\Big]
\nonumber\\
&\qquad
+\frac{2}{1-\nu}\, \epsilon_{rkl}\bigg(
\frac{\delta_{ij}\,R_k+\delta_{ik}\, R_j+\delta_{jk}\, R_i}{R^3}\,
\Big[1-\frac{6\ell^2}{R^2}\,
\Big(1-\e^{-R/\ell}\Big)
+\Big(2+\frac{6\ell}{R}\Big)\,\e^{-R/\ell}\Big]\nonumber\\
&\qquad\quad
-\frac{3 R_iR_j R_k}{R^5}\,
\Big[1-\frac{10\ell^2}{R^2}\,
\Big(1-\e^{-R/\ell}\Big)
+\Big(4+\frac{10\ell}{R}+\frac{2R}{3\ell}\Big)\,\e^{-R/\ell}\Big]\bigg)
\bigg] d l'_r\, .
\end{align}
It can be seen that the Peach-Koehler stress formula~(\ref{T-grad2}) 
is similar to, but more complicated than, the Biot-Savart law~(\ref{B-BS}).

The solution of Eq.~(\ref{u-L}) is the following convolution integral
\begin{align}
\label{u-Mura0}
u_i=-C_{jkln} \pd_k G_{ij}* \beta_{ln}^{\TP,0} .
\end{align}
Using the differentiation rule of a convolution~\citep{Wl}
and Eqs.~(\ref{B-H}) and (\ref{LG-H}), we find the identity
\begin{align}
\label{u-Mura0-1}
u_i=-C_{jkln} \pd_k G_{ij}* L \beta_{ln}^{\TP} 
=-C_{jkln} \pd_k L G_{ij}* \beta_{ln}^{\TP} 
=-C_{jkln} \pd_k G^0_{ij}* \beta_{ln}^{\TP} \,.
\end{align}
This reflects again the relation between the four 
pdes~(\ref{u-L0-RA}), (\ref{u-L0-RA0}), (\ref{u-L0-RA0-1}), and (\ref{u-L0-RA0-2}).
Using Eq.~(\ref{BP-grad0}), it can be represented as a double convolution
\begin{align}
\label{u-Mura0-2}
u_i=-C_{jkln} \pd_k G^0_{ij}*G* \beta_{ln}^{\TP,0}=u_i^0*G \,,
\end{align}
where $u^0_i=-C_{jkln} \pd_k G^0_{ij}* \beta_{ln}^{\TP,0}$.  
Finally, using the relation~(\ref{GT-conv}), Eq.~(\ref{u-Mura0}) is recovered from
Eq.~(\ref{u-Mura0-2}).
Due to the Green function~$G$, Eq.~(\ref{u-Mura0-2}) is the regularization of
the ``classical'' Burgers equation.
If we substitute Eqs.~(\ref{C}) and (\ref{G}) into Eq.~(\ref{u-Mura0}) 
and rearrange terms, we find the displacement field 
in terms of the plastic distortion $\beta_{ln}^{\TP,0}$
\begin{align}
\label{u-Burgers-0}
u_i(\Bx)=-\frac{1}{8\pi}\int_V 
\Big[\big(\delta_{il}\pd_n +\delta_{in}\pd_l -\delta_{ln}\pd_i\big)\Delta
+\frac{1}{1-\nu}\big(\delta_{ln}\Delta - \pd_l\pd_n\big)\pd_i \Big] 
A(R)\, \beta_{ln}^{\TP,0}(\Bx')\, d V'\,.
\end{align}
In the limit $\ell\rightarrow 0$, Eq.~(\ref{u-Burgers-0}) tends
to the classical result given by~\citet{deWit3}.

Substitution of the classical plastic distortion 
of a dislocation loop~(\ref{B0}) into Eq.~(\ref{u-Mura0})
gives the modified Volterra formula
valid in gradient elasticity 
\begin{align}
\label{u-Mura}
u_i(\Bx)=\int_S b_l C_{jkln} \pd_k G_{ij}(R)\, d S'_n\, .
\end{align}
On the other hand,
substituting Eqs.~(\ref{C}) and (\ref{G}) into Eq.~(\ref{u-Mura}), rearranging
terms and using the Stokes theorem,
the key-formula for the non-singular displacement vector 
in gradient elasticity is obtained
\begin{align}
\label{u-Burger-grad}
u_i(\Bx) = -\frac{b_i}{4\pi}\, \Omega(\Bx)
+\frac{b_l\epsilon_{klj}}{8\pi}\, \oint_C
\bigg\{\delta_{ij} \Delta -\frac{1}{1-\nu}\, \pd_i \pd_j   
\bigg\}\, A(R)\,  d l'_k\, ,
\end{align}                  
where the solid angle valid in gradient elasticity is defined by
\begin{align}
\label{Omega}
\Omega(\Bx)=-\frac{1}{2}\, \int_S \Delta\pd_j A(R)\, d S'_j
=\int_S\frac{R_j}{R^3}\Big(1-\Big(1+\frac{R}{\ell}\Big)\e^{-R/\ell}\Big)\, d S'_j
\,.
\end{align}
Eq.~(\ref{u-Burger-grad}) is the Burgers formula valid 
in the framework of gradient elasticity of Helmholtz type.                         
Eq.~(\ref{Omega}) is non-singular and depends on the 
length scale $\ell$.
The solid angle~(\ref{Omega}) 
valid in gradient elasticity can also be transformed into a line 
integral~\citep{Lazar13b}.
Carrying out the differentiations in Eq.~(\ref{u-Burger-grad})
by the help of Eqs.~(\ref{Aij}) and (\ref{Aii}), we obtain 
the explicit gradient elasticity version of the Burgers formula
\begin{align}
\label{u-Burger-grad-2}
u_i(\Bx) = &-\frac{b_i}{4\pi}\, \Omega(\Bx)
-\frac{b_l}{4\pi} \,\oint_C \epsilon_{ilk}\,
\frac{1}{R}\,\Big(1-\e^{-R/\ell}\Big)\, d l'_k
\nonumber\\
&-\frac{b_l}{8\pi (1-\nu)}
\oint_C \epsilon_{ljk}\bigg[
\frac{\delta_{ij}}{R}\Big(1-\frac{2\ell^2}{R^2}\,
\Big(1-\e^{-R/\ell}\Big)+\frac{2\ell}{R}\,\e^{-R/\ell}\Big)\nonumber\\
&\qquad\qquad\qquad
-\frac{R_{i}R_j}{R^3}\Big(1-\frac{6\ell^2}{R^2}\,
\Big(1-\e^{-R/\ell}\Big)
+\Big(2+\frac{6\ell}{R}\Big)\,\e^{-R/\ell}\Big)\bigg]\,  d l'_k\, .
\end{align}    

\subsubsection{Peach-Koehler force between two dislocation loops}

Now, we analyze the Peach-Koehler force in gradient elasticity.
Using the Eshelby stress tensor of gradient elasticity (e.g.,~\citep{LK07})
\begin{align}
\label{P-GE}
P_{kj}=W \delta_{jk} 
-\big(\sigma_{ij}-\pd_l \tau_{ijl}\big) \beta_{ik}-\tau_{ilj}\pd_l \beta_{ik}\,,
\end{align}
the corresponding Peach-Koehler force is obtained
\begin{align}
\int_V \pd_j P_{kj}\, d V=F_k^{\rm{PK}}\,.
\end{align}
The Peach-Koehler force, valid in gradient elasticity of Helmholtz type, 
was originally calculated by \citet{LK07} 
\begin{align}
\label{PK}
F_k^{\rm{PK}}&=\int_V\epsilon_{kjl}\big\{\sigma_{ij}\alpha_{il}+
\tau_{ijm}\, \pd_m\alpha_{il}\big\}\,d V
\nonumber\\
&=\int_V\epsilon_{kjl}\big\{\sigma_{ij}\alpha_{il}+
\ell^2 (\pd_m \sigma_{ij})(\pd_m\alpha_{il})\big\}\, d V
\nonumber\\
&=\int_V\epsilon_{kjl}\big\{\sigma_{ij} L\alpha_{il}+
\ell^2 \pd_m (\sigma_{ij} \pd_m\alpha_{il})\big\}\, d V
\nonumber\\
&=\int_V\epsilon_{kjl} \sigma_{ij}\alpha^0_{il}\, d V\,.
\end{align}
From the third to the fourth line, we used Eq.~(\ref{A-H}) and neglected
the div-term (surface term) at infinity.
%%%It is noted that the structure of the Peach-Koehler force~(\ref{PK}) 
%%%is similar and analogous to the structure 
%%%of the Lorentz force density~(\ref{LF}).

If we substitute Eq.~(\ref{A0}) into Eq.~(\ref{PK}), we find for the
Peach-Koehler force
\begin{align}
\label{PK1}
F_k^{\rm{PK}}=\oint_C\epsilon_{kjm} b_i\sigma_{ij}\, d l'_m\,,
\end{align}
which is the force produced by an ``external'' stress acting on a dislocation 
loop $C$.
Moreover, substituting Eqs.~(\ref{A0}) and (\ref{T-grad}) into Eq.~(\ref{PK}) 
and then integration in $V$,
we obtain the  Peach-Koehler force between the dislocation loop $C^{(A)}$ 
in the stress field of the dislocation loop $C^{(B)}$:
\begin{align}
\label{PK2}
F_m^{\rm{PK}}
=\frac{\mu\, b^{(A)}_i b^{(B)}_k}{8\pi}\oint_{C^{(A)}}\oint_{C^{(B)}}
\epsilon_{mnj}\Big[\big(\epsilon_{jkl}\delta_{ir}
+\epsilon_{ikl}\delta_{jr}\big)\pd_l \Delta
+\frac{2}{1-\nu}\, \epsilon_{rkl}\big(\pd_i\pd_j-\delta_{ij}\Delta\big)\pd_l
\Big] A(R)\, d l^{(B)}_r\, d l^{(A)}_n ,
\end{align}
where $R=|\Bx^{(A)}-\Bx^{(B)}|$ and using Eq.~(\ref{T-grad2}), we get
\begin{align}
\label{PK3}
F_m^{\rm{PK}}
&=\frac{\mu\, b^{(A)}_i b^{(B)}_l}{8\pi}\oint_{C^{(A)}}\oint_{C^{(B)}}
\epsilon_{mnj}
\bigg[\Big(\epsilon_{jkl}\delta_{ir}
+\epsilon_{ikl}\delta_{jr}
-\frac{2}{1-\nu}\,\epsilon_{rkl}\delta_{ij}\Big)
\frac{2R_k}{R^3}\Big[1-\Big(1+\frac{R}{\ell}\Big)\e^{-R/\ell}\Big]
\nonumber\\
&\qquad
+\frac{2}{1-\nu}\, \epsilon_{rkl}\bigg(
\frac{\delta_{ij}\,R_k+\delta_{ik}\, R_j+\delta_{jk}\, R_i}{R^3}\,
\Big[1-\frac{6\ell^2}{R^2}\,
\Big(1-\e^{-R/\ell}\Big)
+\Big(2+\frac{6\ell}{R}\Big)\,\e^{-R/\ell}\Big]\nonumber\\
&\qquad\quad
-\frac{3 R_iR_j R_k}{R^5}\,
\Big[1-\frac{10\ell^2}{R^2}\,
\Big(1-\e^{-R/\ell}\Big)
+\Big(4+\frac{10\ell}{R}+\frac{2R}{3\ell}\Big)\,\e^{-R/\ell}\Big]\bigg)
\bigg]\, d l^{(B)}_r\, d l^{(A)}_n ,
\end{align}
which is non-singular.
The self-force of a dislocation loop can be found 
from the Peach-Koehler force formula~(\ref{PK3})
by using the same curve for $C^{(A)}$ and $C^{(B)}$ and 
the same Burgers vectors $b_i^{(A)}$ and $b_l^{(A)}$.

\subsubsection{Stress functions and 
the elastic interaction energy between two dislocation loops}

Since the stress tensor $\sigma_{ij}$ is symmetric and has zero divergence for
equilibrium in absence of body forces, 
it can be expressed as the inc of a second-order
stress function tensor $B_{ij}$ as (e.g., \citep{Lardner,Teodosiu})
\begin{align}
\label{T-SF}
\sigma_{ij}=-\epsilon_{ikl}\epsilon_{jmn}\pd_k\pd_m B_{ln}\,.
\end{align}
It can be seen that $B_{ij}$ is a symmetric tensor.
Following \citet{Kroener58}, it is convenient to introduce another
symmetric stress function tensor $\chi_{ij}$ which is defined as
\begin{align}
\label{chi}
\chi_{ij}=\frac{1}{2\mu}\, \Big(\delta_{ik}\delta_{jl}
-\frac{\nu}{1+2\nu}\, \delta_{ij}\delta_{kl}\Big)B_{kl}
\end{align}
with the inverse relation
\begin{align}
\label{B-chi}
B_{ij}&=2\mu\, \Big(\delta_{ik}\delta_{jl}
+\frac{\nu}{1-\nu}\, \delta_{ij}\delta_{kl}\Big)\chi_{kl}\,.
\end{align}
The stress function tensor $\chi_{ij}$ satisfies
the following side condition (``Kr\"oner gauge'')
\begin{align}
\pd_j\chi_{ij}=0=\pd_i\chi_{ij}\,.
\end{align}

The so-called incompatibility tensor $\eta_{ij}$ which is defined in terms 
of the elastic strain tensor~\citep{Kroener58,Kroener81,Teodosiu} 
is given by
\begin{align}
\label{ink}
\eta_{ij}=-\epsilon_{ikl}\epsilon_{jmn}\pd_k\pd_m e_{ln}\,.
\end{align}
In terms of the dislocation density tensor~$\alpha_{ij}$, 
the incompatibility tensor $\eta_{ij}$ 
has the form~\citep{Kroener58,Kroener81,Teodosiu} 
\begin{align}
\label{eta}
\eta_{ij}=-\frac{1}{2}\,
\big(\epsilon_{ikl}\pd_k \alpha_{lj}+\epsilon_{jkl}\pd_k \alpha_{li}\big)\,.
\end{align}

On the other hand, the stress tensor fulfills the  
Beltrami-Michell stress incompatibility condition 
(see, e.g., \citep{Kroener58,Teodosiu})
\begin{align}
\label{BM-CC}
\Delta \sigma_{ij}
+\frac{1}{1+\nu}\, \big(\pd_i\pd_j -\delta_{ij}\Delta\big)\sigma_{kk}
=2\mu\,\eta_{ij}\,.
\end{align}
Multiplying Eq.~(\ref{BM-CC}) by the Helmholtz operator $L$, we obtain
\begin{align}
\label{BM-CCL}
L\Big[\Delta \sigma_{ij}
+\frac{1}{1+\nu}\, \big(\pd_i\pd_j -\delta_{ij}\Delta\big)\sigma_{kk}\Big]
=2\mu\,\eta^0_{ij}\,,
\end{align}
with  
\begin{align}
\label{eta0}
\eta^0_{ij}=-\frac{1}{2}\,
\big(\epsilon_{ikl}\pd_k \alpha^0_{lj}+\epsilon_{jkl}\pd_k \alpha^0_{li}\big)\,,
\end{align}
where we used Eq.~(\ref{A-H}) and 
\begin{align}
\label{eta-L}
L \eta_{ij}=\eta^0_{ij}\,.
\end{align}

Substituting Eq.~(\ref{B-chi}) into Eq.~(\ref{T-SF}), the stress tensor
reads in terms of the stress function tensor $\chi_{ij}$ as
\begin{align}
\label{T-chi}
\sigma_{ij}=
2\mu\Big(\Delta \chi_{ij}
+\frac{1}{1-\nu}\, \big(\pd_i\pd_j -\delta_{ij}\Delta\big)\chi_{kk}\Big)\,.
\end{align}
If we substitute Eq.~(\ref{T-chi}) into Eq.~(\ref{BM-CCL}),
we obtain
\begin{align}
\label{chi-pde}
L \Delta\Delta\, \chi_{ij}= \eta^0_{ij}\,,
\end{align}
which is an inhomogeneous Helmholtz-bi-Laplace equation for
$\chi_{ij}$.
The Green function of the Helmholtz-bi-Laplace equation (pde of 6th-order) 
is defined as
\begin{align}
\label{LDD-G}
L \Delta\Delta G=\delta(\Bx-\Bx')\,.
\end{align}
Comparing Eq.~(\ref{LDD-G}) with Eq.~(\ref{A-LDD}), the Green function can be 
written in terms of the ``regularization function''~(\ref{A}).
Thus, the Green function of the Helmholtz-bi-Laplace equation
is given by
\begin{align}
\label{G-LBH}
G(R)=-\frac{1}{8\pi}\, A(R)
=-\frac{1}{8\pi}\bigg(R+\frac{2\ell^2}{R}\,\Big(1-\e^{-R/\ell}\Big)\bigg)\, .
\end{align} 
Some remarks on the Green function of the Helmholtz-bi-Laplace equation
given by~\citet{Eringen84,Eringen02} in the framework of
nonlocal elasticity of Helmholtz type are following. 
The Green function given by~\citet{Eringen84,Eringen85,Eringen02}
is not the correct one since the second term of Eq.~(\ref{G-LBH}), $2\ell^2/R$, 
is missing in Eringen's expression for $G$ 
(compare with Eq.~(6.13.24) in~\citep{Eringen02}). 
Therefore, Eringen's expression for $G$
does not give the correct Green function of 
the Helmholtz-bi-Laplace equation~(\ref{LDD-G}).
As a consequence, the derived Peach-Koehler stress formula
based on the mistaken expression for $G$
remains still singular.
Moreover, using the correct Green function~(\ref{G-LBH}), 
one can derive the correct Peach-Koehler stress formula in nonlocal elasticity 
of Helmholtz type, which agrees with the Peach-Koehler stress formula~(\ref{T-grad}) 
in gradient elasticity of Helmholtz type.
The Peach-Koehler stress formula~(\ref{T-grad2}) based on the Green function~(\ref{G-LBH})
is not singular.
Thus, the expressions~(\ref{T-grad}) and (\ref{T-grad2}) represent the correct 
Peach-Koehler stress tensor field in the framework of nonlocal elasticity of
Helmholtz type as well.
For gradient elasticity of bi-Helmholtz type~\citep{LMA06a}
and nonlocal elasticity of bi-Helmholtz type~\citep{LMA06b}
the regularization function $A(R)$ and the corresponding Green function $G(R)$
of the bi-Helmholtz-bi-Laplace equation can be found in~\citep{Lazar13}.

The solution of Eq.~(\ref{chi-pde}) for an infinite solid 
may be given by
\begin{align}
\chi_{ij}=G*\eta^0_{ij}\,.
\end{align}
If we substitute Eqs.~(\ref{eta0}) and (\ref{A0}) and calculate the
convolution integral, this gives
\begin{align}
\label{chi-L}
\chi_{ij}=\frac{b_l}{8\pi}\frac{1}{2}\,
\Big(\epsilon_{ikl}\oint_C \pd_k A(R)\, d l'_j
+\epsilon_{jkl}\oint_C \pd_k A(R)\, d l'_i\Big)\,,
\end{align}
where we have used the Green-Gauss theorem and set a surface term 
at infinity to zero.
The trace term of the stress function tensor reads now
\begin{align}
\label{chi-L-Tr}
\chi_{ii}=\frac{b_l}{8\pi}\,
\epsilon_{ikl}\oint_C \pd_k A(R)\, d l'_i\,.
\end{align}
Upon substituting Eqs.~(\ref{chi-L}) and (\ref{chi-L-Tr}) into Eq.~(\ref{T-chi}), 
the Peach-Koehler stress formula~(\ref{T-grad}) is obtained.

Now, we turn to the interaction energy.
According to the theory of gradient elasticity, the interaction 
energy can be written as
\begin{align}
\label{W-int1}
W^{(AB)}&=\int_V \Big(\sigma^{(B)}_{ij} e^{(A)}_{ij}
+\ell^2\pd_k \sigma^{(B)}_{ij}\pd_k  e^{(A)}_{ij}\Big)
\, d V
=\int_V \sigma^{(B)}_{ij}\,L  e^{(A)}_{ij}\, d V\,,
\end{align}
where we have used again the Green-Gauss theorem and set the surface term 
at infinity to zero.
By partial integration and using Eqs.~(\ref{T-chi}), (\ref{eta}) and (\ref{eta-L}),
Eq.~(\ref{W-int1}) can be transformed into
\begin{align}
\label{W-int2}
W^{(AB)}&=-\int_V \big(\epsilon_{ikl}\epsilon_{jmn}\, \pd_k\pd_m B^{(B)}_{ln}\big)\,L e^{(A)}_{ij}\, d V
\nonumber\\
&=-\int_V B^{(B)}_{ln}\,\big(\epsilon_{ikl}\epsilon_{jmn}\, \pd_k\pd_m \,L e^{(A)}_{ij}\big)\, d V
\nonumber\\
&=\int_V B^{(B)}_{ij}\big(L \eta^{(A)}_{ij}\big)\, d V
\nonumber\\
&=\int_V B^{(B)}_{ij}\,\eta^{0,(A)}_{ij}\, d V\,.
\end{align}
Now, substituting Eqs.~(\ref{B-chi}), (\ref{eta0}) and (\ref{A0}) into
Eq.~(\ref{W-int2}), we obtain after the volume integration 
\begin{align}
\label{W-int3}
W^{(AB)}&=2\mu \int_V \Big(\chi^{(B)}_{ij}
+\frac{\nu}{1-\nu}\, \delta_{ij}\chi^{(B)}_{kk}\Big) \eta^{0,(A)}_{ij}\, d V
\nonumber\\
&=2\mu\, \epsilon_{ikl}\, b^{(A)}_l \oint_{C^{(A)}} \pd_k \Big(\chi^{(B)}_{ij}
+\frac{\nu}{1-\nu}\, \delta_{ij}\,\chi^{(B)}_{pp}\Big) d l^{(A)}_j\,.
\end{align}
Eq.~(\ref{W-int3}) represents the energy of a dislocation
line ``running'' along the curve $C^{(A)}$ with Burgers vector 
$b_l^{(A)}$ interacting with 
a field whose stress function is given by $\chi^{(B)}_{ij}$. 
If we substitute Eqs.~(\ref{chi-L}) and (\ref{chi-L-Tr}) into Eq.~(\ref{W-int3}), 
we find the mutual interaction energy between two closed dislocation loops 
\begin{align}
\label{WAB1}
W^{(AB)}
=\frac{\mu}{8\pi}\, b^{(A)}_i b^{(B)}_j
\oint_{C^{(A)}}\oint_{C^{(B)}}
\epsilon_{ikl}\epsilon_{jmn}\, \pd_k\pd_m\, A(R)
\Big(d l^{(B)}_l\, d l^{(A)}_n +\delta_{ln}\,d l^{(B)}_p\, d l^{(A)}_p
+\frac{2\nu}{1-\nu}\,  d l^{(B)}_n\, d l^{(A)}_l \Big)\,.
\end{align}
In the limit $\ell\rightarrow 0$, 
the form of the interaction energy given by~\citet{Kroener58,Kroener81} 
(see also~\citep{deWit60,Lardner,Kleinert}) is obtained.
Eq.~(\ref{WAB1}) may be re-written as
\begin{align}
\label{WAB2}
W^{(AB)}
=b^{(A)}_i b^{(B)}_j M^{(AB)}_{ij}
\end{align}
with the so-called ``dislocation mutual inductance'' tensor,
which is in general asymmetric,  
\begin{align}
\label{MAB1-disl}
M^{(AB)}_{ij}
=\frac{\mu}{8\pi}
\oint_{C^{(A)}}\oint_{C^{(B)}}
\epsilon_{ikl}\epsilon_{jmn}\, \pd_k\pd_m\, A(R)
\Big(d l^{(B)}_l\, d l^{(A)}_n +\delta_{ln}\,d l^{(B)}_p\, d l^{(A)}_p
+\frac{2\nu}{1-\nu}\,  d l^{(B)}_n\, d l^{(A)}_l \Big)\,.
\end{align}

On the other hand, Eq.~(\ref{WAB1}) can be simplified to
\begin{align}
\label{WAB3}
W^{(AB)}
&=-\frac{\mu}{8\pi}\, b^{(A)}_i b^{(B)}_j
\oint_{C^{(A)}}\oint_{C^{(B)}}\Big[
\Delta\, A(R)
\Big(d l^{(B)}_j\, d l^{(A)}_i +\frac{2\nu}{1-\nu}\, d l^{(B)}_i\, d l^{(A)}_j
\Big)\nonumber\\
&\hspace{5cm}
+\frac{2}{1-\nu}\, \big(\pd_i\pd_j -\delta_{ij}\,\Delta\big)\, A(R)
\,  d l^{(B)}_k\, d l^{(A)}_k \Big]\,
\end{align}
and the corresponding ``dislocation mutual inductance'' tensor is given by
\begin{align}
\label{MAB3}
M^{(AB)}_{ij}
&=-\frac{\mu}{8\pi}
\oint_{C^{(A)}}\oint_{C^{(B)}}\Big[
\Delta\, A(R)
\Big(d l^{(B)}_j\, d l^{(A)}_i +\frac{2\nu}{1-\nu}\, d l^{(B)}_i\, d l^{(A)}_j
\Big)\nonumber\\
&\hspace{5cm}
+\frac{2}{1-\nu}\, \big(\pd_i\pd_j -\delta_{ij}\,\Delta\big)\, A(R)
\,  d l^{(B)}_k\, d l^{(A)}_k \Big]\,.
\end{align}
In the limit $\ell\rightarrow 0$, 
the form of the interaction energy given by~\citet{deWit60,deWit67} is recovered.
By use of the Eqs.~(\ref{Aij}) and ~(\ref{Aii}), Eq.~(\ref{WAB3}) reads explicitly
\begin{align}
\label{WAB4}
W^{(AB)}
&=-\frac{\mu}{8\pi}\, b^{(A)}_i b^{(B)}_j
\oint_{C^{(A)}}\oint_{C^{(B)}}\bigg[
\frac{2}{R}\Big(1-\e^{-R/\ell}\Big)
\Big(d l^{(B)}_j\, d l^{(A)}_i +\frac{2\nu}{1-\nu}\, d l^{(B)}_i\, d l^{(A)}_j
\Big)\nonumber\\
&\hspace{2cm}
+\frac{2}{1-\nu}
\bigg(\frac{\delta_{ij}}{R}\Big[1-\frac{2\ell^2}{R^2}\,
\Big(1-\e^{-R/\ell}\Big)+\frac{2\ell}{R}\,\e^{-R/\ell}\Big]
-\frac{2\delta_{ij}}{R}\Big(1-\e^{-R/\ell}\Big)\nonumber\\
&\hspace{2cm}
-\frac{R_{i}R_j}{R^3}\Big[1-\frac{6\ell^2}{R^2}\,
\Big(1-\e^{-R/\ell}\Big)
+\Big(2+\frac{6\ell}{R}\Big)\,\e^{-R/\ell}\Big]\bigg)
\,  d l^{(B)}_k\, d l^{(A)}_k \bigg]\,,
\end{align}
where it can be easily seen that the interaction energy is non-singular.
The corresponding ``dislocation mutual inductance'' tensor is
\begin{align}
\label{MAB1-disl2}
M^{(AB)}_{ij}
&=-\frac{\mu}{8\pi}
\oint_{C^{(A)}}\oint_{C^{(B)}}\bigg[
\frac{2}{R}\Big(1-\e^{-R/\ell}\Big)
\Big(d l^{(B)}_j\, d l^{(A)}_i +\frac{2\nu}{1-\nu}\, d l^{(B)}_i\, d l^{(A)}_j
\Big)\nonumber\\
&\hspace{2cm}
+\frac{2}{1-\nu}
\bigg(\frac{\delta_{ij}}{R}\Big[1-\frac{2\ell^2}{R^2}\,
\Big(1-\e^{-R/\ell}\Big)+\frac{2\ell}{R}\,\e^{-R/\ell}\Big]
-\frac{2\delta_{ij}}{R}\Big(1-\e^{-R/\ell}\Big)\nonumber\\
&\hspace{2cm}
-\frac{R_{i}R_j}{R^3}\Big[1-\frac{6\ell^2}{R^2}\,
\Big(1-\e^{-R/\ell}\Big)
+\Big(2+\frac{6\ell}{R}\Big)\,\e^{-R/\ell}\Big]\bigg)
\,  d l^{(B)}_k\, d l^{(A)}_k \bigg]\,.
\end{align}
The self-energy of a dislocation loop can be found by using the same
curve for $C^{(A)}$ and $C^{(B)}$, so that
$M_{ij}^{(AA)}$ becomes the tensor of ``dislocation self-inductance''.
Inserting a factor $\frac{1}{2}$, we find:
$W^{(AA)}=\frac{1}{2}\, b_i^{(A)}b_j^{(A)} M_{ij}^{(AA)}$.

Thus, in subsection~\ref{DL} 
we have seen that the Burgers, Mura, Peach-Koehler stress, 
Peach-Koehler force and the mutual interaction energy formulas 
are non-singular in the
framework of gradient elasticity theory of Helmholtz type.
Finally, one can observe that 
all these dislocation key-formulas can be obtained from their classical 
counterparts by means of the substitution:
$R\rightarrow A(R)$.
On the other hand, substituting the decomposition of the 
``regularization function''~(\ref{A-bi}) into the dislocation key-formulas,
the classical term and the gradient term of the dislocation key-formulas 
are easily obtained 
corresponding to the classical term $A^0$ and the gradient term $A^1$.
Gradient elasticity is a theory with dislocation core regularization. 
This is not only necessary for the explanation of physical core effects, but 
also for the elimination of singularities in a physically well founded manner 
in numerical simulations.
In the limit $\ell\rightarrow 0$, the classical singular dislocation key-formulas 
are obtained from the non-singular ones (see, e.g.,~\citep{HL,Lardner,Teodosiu,Li}).
These results may be used in computer simulations of discrete dislocation dynamics
and in the numerics as fast numerical sums of the relevant elastic fields
as they are used for the classical equations (e.g.,~\citep{Sun,Ghoniem02}).
One of the main limitations of current dislocation dynamics models is the 
inability to resolve
dislocation interactions in close range without ad-hoc or more sophisticated
regularization strategies. 
The regularization offered here by the gradient theory is 
particularly advantageous for dislocation dynamics simulations.
The 3D non-singular dislocation fields can be implemented 
in 3D dislocation dynamics codes~\citep{Po13}.
This can represent the breakthrough of gradient elasticity in the modeling
of dislocation dynamics without singularities. 
Such a dislocation dynamics without 
singularities offers the promise of predicting the dislocation microstructure
evolution from fundamental principles and based on sound physical grounds.
Therefore, a dislocation-based plasticity theory can be based on gradient 
elasticity theory of non-singular dislocations.

\section{Conclusion}
In this paper, the gradient theory of magnetostatics has been presented as part of
the Bopp-Podolsky theory in order to show how gradient theories are used in physics.
We have investigated an electric current loop and the Biot-Savart law. 
Using the theory of gradient magnetostatics, we found
non-singular solutions for all relevant fields in analogy to 
the ``classical'' singular solutions of magnetostatics.
Also, the so-called ``Bifield'' ansatz has been discussed in this framework.

In the main part of the paper, 
the theory of gradient elasticity of Helmholtz type has been presented and
investigated.
Many analogies and similarities between gradient magnetostatics and gradient
elasticity of Helmholtz type have been pointed out.
Furthermore, non-singular dislocation key-formulas have been 
presented in the framework of gradient elasticity. 
The technique of Green functions has been used. 
A ``Bifield'' ansatz has been used 
and the ``Ru-Aifantis theorem'' has been generalized to the problem 
of dislocations in gradient elasticity of Helmholtz type.
From the field theoretical point of view, 
the theory of gradient elasticity is similar to, but more complicated than,
the theory of gradient magnetostatics. 
The elastic distortion, plastic 
distortion, stress, displacement, and dislocation density 
of a closed dislocation loop were calculated using the theory of 
gradient elasticity of Helmholtz type. 
Such a generalized continuum theory allows
dislocation core spreading in a straightforward way.
In the theory of gradient elasticity all formulas are closed and self-consistent.
It should be emphasized that 
the Green function, $G$, of the Helmholtz equation plays the role of the 
regularization function in gradient elasticity of Helmholtz type.
In addition, we have found two important basic-results for the theory 
of gradient elasticity of Helmholtz type.
First, we have shown that the tensor,
$\sigma_{ij}^0=\sigma_{ij}-\pd_k\tau_{ijk}$, 
is identical with
the classical stress tensor and, therefore, there is no need to call such a
tensor as total stress tensor. 
Second, using the theory of generalized functions,
we have shown that the Cauchy stress tensor of gradient elasticity 
$\sigma_{ij}$ is self-equilibrated, $\pd_j \sigma_{ij}=0$.

The obtained dislocation key-formulas can be used in computer simulations and numerics of 
discrete dislocation dynamics
of arbitrary 3D dislocation configurations.
They can be implemented in dislocation dynamics codes (finite element
implementation, technique of fast numerical sums, 
method of parametric dislocation dynamics), 
and compared to atomistic models (e.g.,~\citep{Sun,Li}). 
Thus, the obtained 
non-singular dislocation key-formulas serve the basis of a
non-singular discrete dislocation dynamics.

\section*{Acknowledgements}
The author gratefully acknowledges 
Dr.~Eleni Agiasofitou for many fruitful discussions and constructive remarks, 
which significantly influenced this work.
The author acknowledges the grants from the 
Deutsche Forschungsgemeinschaft 
(Grant Nos. La1974/2-1, La1974/2-2, La1974/3-1).

\begin{appendix}
\section{Derivatives of the 
``regularization function'' $A$}
\label{appendixA}
\setcounter{equation}{0}
\renewcommand{\theequation}{\thesection.\arabic{equation}}

In this appendix, the relevant derivatives of the 
``regularization function'' $A$ are given.
For gradient elasticity of Helmholtz type, the elementary function $A$ is given by
\begin{align}
\label{A2}
A=R+\frac{2\ell^2}{R}\,
\Big(1-\e^{-R/\ell}\Big)\, .
\end{align} 
The higher-order derivatives of $A$ are given by the following set of equations
\begin{align}
\label{Ai}
\pd_i A_{}=\frac{R_i}{R}\Big[1-\frac{2\ell^2}{R^2}\,
\Big(1-\e^{-R/\ell}\Big)+\frac{2\ell}{R}\,\e^{-R/\ell}\Big]
\, ,
\end{align} 
where $R_i=x_i-x'_i$,
\begin{align}
\label{Aij}
\pd_j\pd_i A_{}=\frac{\delta_{ij}}{R}\Big[1-\frac{2\ell^2}{R^2}\,
\Big(1-\e^{-R/\ell}\Big)+\frac{2\ell}{R}\,\e^{-R/\ell}\Big]
-\frac{R_{i}R_j}{R^3}\Big[1-\frac{6\ell^2}{R^2}\,
\Big(1-\e^{-R/\ell}\Big)
+\Big(2+\frac{6\ell}{R}\Big)\,\e^{-R/\ell}\Big]
\, ,
\end{align} 
\begin{align}
\label{Aii}
\pd_i \pd_i A_{}=\frac{2}{R}\Big(1-\e^{-R/\ell}\Big)\,,
\end{align} 
\begin{align}
\label{Aijk}
\pd_k\pd_j\pd_i A_{}=&-\frac{\delta_{ij}\,R_k+\delta_{ik}\, R_j+\delta_{jk}\, R_i}{R^3}\,
\Big[1-\frac{6\ell^2}{R^2}\,
\Big(1-\e^{-R/\ell}\Big)
+\Big(2+\frac{6\ell}{R}\Big)\,\e^{-R/\ell}\Big]\nonumber\\
&+\frac{3 R_iR_j R_k}{R^5}\,
\Big[1-\frac{10\ell^2}{R^2}\,
\Big(1-\e^{-R/\ell}\Big)
+\Big(4+\frac{10\ell}{R}+\frac{2R}{3\ell}\Big)\,\e^{-R/\ell}\Big]
\, 
\end{align} 
and
\begin{align}
\label{Aiik}
\pd_k \pd_i\pd_i A_{}&=-\frac{2R_k}{R^3}\Big(1-\Big(1+\frac{R}{\ell}\Big)\e^{-R/\ell}\Big)
\,.
\end{align}
The expressions~(\ref{A2})--(\ref{Aiik}) are non-singular.

\section{Boundary conditions in gradient elasticity}
\label{appendixB}
\setcounter{equation}{0}
The general form of the boundary conditions (BCs) corresponding to Eq.~(\ref{EC})
in gradient elasticity 
reads (see, e.g.,~\citep{Mindlin68,Jaunzemis,Gao07})
\begin{align}
\label{BC}
\left.
\begin{array}{r}
\displaystyle{\big(\sigma_{ij}-\pd_k\tau_{ijk}\big)n_j-\pd_j\big(\tau_{ijk} n_k\big)
+n_j\pd_l\big(\tau_{ijk} n_k n_l\big)=\bar{t}_i}\\
\displaystyle{\tau_{ijk}n_jn_k=\bar{q}_i}\\
\end{array}
\right\}
\qquad\text{on}\qquad \pd\Omega\,,
\end{align}
where ${t}_i$ and ${q}_i$ are the Cauchy traction vector and the 
double stress traction vector, respectively. 
Moreover, $\pd\Omega$ is the smooth boundary surface of the domain $\Omega$ 
occupied by the body satisfying the Euler-Lagrange equation~(\ref{EC}),
$n_i$ denotes the unit outward-directed vector normal to the boundary $\pd\Omega$ ,
and the overhead bar represents the prescribed value.
Using the constitutive equation~(\ref{CR2}) and Eq.~(\ref{RA2}), the
BCs~(\ref{BC}) simplify to the form
\begin{align}
\label{BC2}
\left.
\begin{array}{r}
\displaystyle{\sigma^0_{ij}n_j-\ell^2 \pd_j\big(  n_k\pd_k \sigma_{ij}\big)
+\ell^2 n_j \pd_l\big( n_l  n_k \pd_k \sigma_{ij}\big)=\bar{t}_i}\\
\displaystyle{\ell^2 n_j n_k \pd_k\sigma_{ij}=\bar{q}_i}\\
\end{array}
\right\}
\qquad\text{on}\qquad \pd\Omega\,.
\end{align}
In addition, BC~(\ref{BC2}a) can be written as~\citep{GM10a}
\begin{align}
\label{BC2a}
\sigma^0_{ij}n_j
-\ell^2 \big[  (\pd_j n_k) \pd_k \sigma_{ij}
+n_k \pd_k \pd_j\sigma_{ij}\big]
+\ell^2 n_j\big[ (\pd_l n_l)  n_k \pd_k \sigma_{ij}
+n_l (\pd_l n_k)  \pd_k \sigma_{ij}
+n_l n_k \pd_l \pd_k \sigma_{ij}\big]=\bar{t}_i\,.
\end{align}
If $n_i$ is constant, then the BC~(\ref{BC2a}) simplifies to
\begin{align}
\label{BC2b}
\sigma^0_{ij}n_j
-\ell^2 \big[n_k \pd_k \pd_j\sigma_{ij}
-n_j n_l n_k \pd_l \pd_k \sigma_{ij}\big]=\bar{t}_i\,.
\end{align}
Using Eq.~(\ref{DivT}), the BC~(\ref{BC2b}) reduces to
\begin{align}
\label{BC2c}
\sigma^0_{ij}n_j
+\ell^2 n_j n_l n_k \pd_l \pd_k \sigma_{ij}=\bar{t}_i\,.
\end{align}
In the limit $\ell\rightarrow 0$, the BCs~(\ref{BC2}) reduce to the classical
one: $\sigma^0_{ij} n_j=\bar{t}_i$.

Using the ``Bifield'' ansatz~(\ref{t-bi}), the BCs~(\ref{BC2}) 
can be decomposed into the classical part for $\sigma^0_{ij}$ 
and a gradient part for $\sigma^1_{ij}$ (see also~\citep{Ara}).
In this manner, the classical part of the BCs corresponding to 
the classical equilibrium condition~(\ref{FE-Bi-0}) reads
\begin{align}
\label{BC2-0}
\displaystyle{\sigma^0_{ij}n_j=\bar{t}_i}
\qquad\text{on}\qquad \pd\Omega\,
\end{align}
and the gradient part of the BCs corresponding to the field equation~(\ref{FE-Bi-1})
is given by
\begin{align}
\label{BC2-1}
\left.
\begin{array}{c}
\displaystyle{-\ell^2 \pd_j(  n_k\pd_k \sigma^1_{ij})
+\ell^2 n_j \pd_l( n_l  n_k \pd_k \sigma^1_{ij})=
\ell^2 \pd_j(  n_k\pd_k \sigma^0_{ij})
-\ell^2 n_j \pd_l ( n_l  n_k \pd_k \sigma^0_{ij})}\\
\qquad\quad
\displaystyle{\ell^2 n_j n_k \pd_k\sigma^1_{ij}=\bar{q}_i
-\ell^2 n_j n_k \pd_k\sigma^0_{ij}}\\
\end{array}
\right\}
\quad\text{on}\quad \pd\Omega\,.
\end{align}
It can be seen in Eq.~(\ref{BC2-1}) that the classical stress $\sigma^0_{ij}$ acts 
also as traction for the gradient part $\sigma^1_{ij}$.

If $n_i$ is constant, $\pd_j \sigma_{ij}^0=0$,  $\pd_j \sigma_{ij}^1=0$ are
fulfilled and using the BC~(\ref{BC2-0}), we find
\begin{align}
\label{BC2-2}
\left.
\begin{array}{c}
\displaystyle{
\ell^2 n_j n_l  n_k \pd_l \pd_k \sigma^1_{ij}=
-\ell^2 n_l  n_k \pd_l \pd_k \bar{t}_{i}}\\
\qquad\quad
\displaystyle{\ell^2 n_j n_k \pd_k\sigma^1_{ij}=\bar{q}_i
-\ell^2 n_k \pd_k \bar{t}_{i}}\\
\end{array}
\right\}
\quad\text{on}\quad \pd\Omega\,.
\end{align}
In addition, if the Cauchy traction $\bar{t}_{i}$ is constant, 
then the BCs~(\ref{BC2-2}) simplify to
\begin{align}
\label{BC2-3}
\displaystyle{\ell^2 n_j n_k \pd_k\sigma^1_{ij}=\bar{q}_i}
\qquad\text{on}\qquad \pd\Omega\,
\end{align}
and 
\begin{align}
\label{BC2-4}
\displaystyle{\ell^2 n_j n_l n_k \pd_l \pd_k\sigma^1_{ij}=
n_l\pd_l \bar{q}_i=0}
\qquad\text{on}\qquad \pd\Omega\,.
\end{align}
Eq.~(\ref{BC2-4}) is fulfilled if the double traction $\bar{q}_i$ is constant.
Thus, for constant vector normal, constant Cauchy traction,
constant double stress traction and using the ``Bifield'' ansatz the BCs 
of gradient elasticity simplify to the expressions~(\ref{BC2-0}) and (\ref{BC2-3}).
The BC~(\ref{BC2-0}) relates the Cauchy traction to the classical
Cauchy stress tensor and the  BC~(\ref{BC2-3}) connects the double stress
traction with the gradient part of the Cauchy stress tensor.

\end{appendix}

\end{document}